\documentclass[a4paper,11pt,superscriptaddress,preprintnumbers,floatfix]{article}
\usepackage{jheppub} 
\usepackage{subcaption}

\usepackage[usenames, dvipsnames]{color}
\usepackage[utf8]{inputenc}
\usepackage{amssymb,amsthm,amsmath}

\usepackage{multirow}
\usepackage{slashed}
\usepackage[titletoc,title]{appendix}
\usepackage{array}
\usepackage{float} 

\hypersetup{
  colorlinks,
  citecolor=black,
  filecolor=black,
  linkcolor=black,
  urlcolor=black
}
\usepackage{braket}
\usepackage{bm}
\usepackage[export]{adjustbox}
\usepackage{xcolor}

\usepackage{cleveref}
\crefname{section}{Sec.}{Secs.}
\crefname{figure}{Fig.}{Figs.}
\crefname{appendix}{Appendix}{Appendices}
\crefname{equation}{Eq.}{Eqs.}
\crefname{table}{Table}{Tables}
\Crefname{section}{Section}{Sections}
\Crefname{figure}{Figure}{Figures}
\Crefname{appendix}{Appendix}{Appendices}
\Crefname{equation}{Equation}{Equations}
\Crefname{table}{Table}{Tables}

\usepackage{quantikz}  

\colorlet{darkgreen}{green!50!black}
\colorlet{brightyellow}{yellow!75!red}
\colorlet{orange}{red!50!yellow}
\colorlet{darkgray}{gray!50!black}
\colorlet{darkred}{red!50!black}
\colorlet{darkblue}{blue!50!black}

\usepackage{subcaption}
\DeclareMathOperator{\Tr}{Tr}

\def\cT{{\cal T}}
\newcommand*\diff{\mathop{}\!\mathrm{d}}

\newcommand{\ind}{{\beta}} 
\newcommand{\pd}{{\vphantom{\dagger}}}
\newcommand{\GeV}{{{\,}\textrm{GeV}}}

\title{Real-time quantum simulation of hadron production in deep inelastic scattering}

\author[a,*]{Xianghui Cao,\note[*]{Authors are listed in alphabetical order.}}
\emailAdd{xianghuicao@mail.ustc.edu.cn}
\affiliation[a]{Department of Modern Physics, University of Science and Technology of China, Hefei 230026, China }

\author[b,*]{Meijian Li,}
\emailAdd{meijianli@hust.edu.cn}
\affiliation[b]{School of Physics, Huazhong University of Science and Technology, Wuhan 430074, China}

\author[a,c,*]{Yang Li,}
\emailAdd{leeyoung1987@ustc.edu.cn}
\affiliation[c]{Anhui Center for Fundamental Sciences in Theoretical Physics, Hefei, 230026, China }

\author[d,*]{and Wenyang Qian}
\emailAdd{wqian@ccnu.edu.cn}
\affiliation[d]{Institute of Particle Physics and Key Laboratory of Quark and Lepton Physics (MOE),
Central China Normal University, Wuhan, 430079, Hubei, China}

\abstract{
We develop a quantum simulation framework for meson production in deep inelastic scattering based on light-front QCD Hamiltonian formalism.
A quark-antiquark dipole, prepared either as a single-momentum state or as the $q\bar q$ component of a virtual photon, is evolved in real time through the color field of a nuclear target via Trotterized unitary evolution.
We consider both proton and gold nuclear targets, each characterized by their respective saturation scale. 
From the evolved state we extract the transverse momentum broadening, the total and elastic cross sections, and the $J/\psi$ production rate by projecting onto the $J/\psi$ light-front wavefunction.
We then extract the nuclear modification factor $R_A$ at different photon virtuality $Q^2$ and analyze nuclear modification effect.
Results of classical simulations show good agreement with analytical eikonal predictions across a broad range of parameters. 
For the single-momentum dipole, quantum simulation is demonstrated on a quantum circuit emulator with a smaller lattice size, and results are consistent with the classical benchmark.
This work establishes a first-principles, amplitude-level pathway toward computing exclusive hadron production in deep inelastic scattering, directly relevant to the physics program of the Electron-Ion Collider.
}

\begin{document}
\maketitle
\flushbottom

\section{Introduction}

Understanding how quarks and gluons produced in high-energy collisions evolve into observable hadrons remains one of the central open problems in QCD~\cite{Albino:2008gy,Metz:2016swz}. In deep inelastic scattering (DIS), a virtual photon splits into a quark-antiquark dipole that propagates through the color field of the nuclear target and subsequently evolves into final-state hadrons. In this small-$x$ regime, the target color field is described within the Color Glass Condensate (CGC) effective theory, and the dipole-target interaction is computed from correlators of Wilson lines, which determine the total and elastic dipole cross sections~\cite{Gelis:2010nm,Iancu:2003xm,Dumitru:2002qt}. 
Exclusive vector-meson production, in particular $J/\psi$ production, is a sensitive probe of the dense gluon field and its nuclear modification, and has been studied extensively within dipole and saturation-based approaches~\cite{Kowalski:2006hc,Lappi:2013am,Mantysaari:2022ffw,Mantysaari:2023qsq,Mantysaari:2023xcu,Dai:2026rgl}. The underlying dynamics, however, spans multiple scales and regimes: the initial splitting is perturbative, while the final hadron formation is intrinsically nonperturbative that involves quantum coherence across the entire propagation history. Existing approaches rely on factorization theorems combined with phenomenological fragmentation functions~\cite{Metz:2016swz} or semi-classical generators~\cite{Andersson:1983ia,Sjostrand:2006za} that break quantum coherence, and, together with the eikonal approximation underlying the dipole picture, cannot systematically access the real-time amplitude-level dynamics~\cite{Mueller:1997ik,Kovchegov:1999kx}. A first-principles, real-time, nonperturbative description of the process from dipole formation through hadron production is therefore highly desirable, both for understanding confinement and for predicting observables at facilities such as the Electron-Ion Collider (EIC)~\cite{Accardi:2012qut,AbdulKhalek:2021gbh}.

Quantum computing offers a natural path to overcome these limitations: the ability to track unitary time evolution of entangled many-body states without exponential classical overhead makes quantum computers ideally suited to real-time, amplitude-level QCD dynamics~\cite{DiMeglio:2023nsa,Bauer:2022hpo,Bauer:2023qgm,Fang:2024ple}.
Particularly, simulating high-energy real-time scattering processes on quantum machines or architecture has attracted growing attention in recent years, with proposals and implementations targeting QCD jet modification in medium~\cite{Barata:2021yri,Barata:2022wim,Barata:2023clv,Qian:2024gph,Wu:2024adk,Castro:2025ocx,Barata:2026icn,Yao:2022eqm}, hadronic particle scattering~\cite{Papaefstathiou:2024zsu,Kreshchuk:2023btr,Li:2023kex,Barata:2025rjb,Barata:2025hgx,Chai:2025kbi,Chai:2025qhf,Chai:2023qpq,Gupta:2026tcg,Bennewitz:2025nhz,Davoudi:2025rdv,Schuhmacher:2025ehh,Belyansky:2023rgh,Artiaco:2025qqq}, real-time partonic correlation~\cite{Schneider:2024yub,Banuls:2025wiq,Li:2021kcs,Li:2024nod,Li:2022lyt,Li:2026thq,Zou:2026cfk,Ikeda:2025bjb,Kang:2025xpz,Grieninger:2025mbm,Grieninger:2024cdl}, and non-equilibrium dynamics towards equilibration~\cite{Halimeh:2025vvp,Mueller:2024mmk,Du:2023ewh,Cuntin:2026nlz,deJong:2021wsd,DeJong:2020riy}. A key challenge in all such applications is the design of efficient encodings that map the physical degrees of freedom onto qubits and the construction of structured quantum circuits that implement the time evolution with polynomial gate complexity.

Light-front quantization provides a natural framework for studying the high-energy scattering process~\cite{Dirac:RevModPhys.21.392,Brodsky:1997de}: the light-front QCD Hamiltonian $P^-$ governs real-time scattering directly in Minkowski space using light-front time $x^+$, and hadronic states admit a Fock-space expansion in terms of on-shell partons. 
In recent years, real-time simulation framework has been built based on this formalism and extensively applied to quark jet propagation through gluon background fields in cold nuclear matter and hot dense media~\cite{Li:2020uhl,Li:2021zaw,Li:2023jeh,Li:2025wzq,Avramescu:2026qro}.
This Fock-space structure maps efficiently onto quantum circuits through a direct encoding scheme in which each single-particle mode is assigned to one qubit register~\cite{Qian:2024gph,Kreshchuk:2020dla}, yielding qubit counts that scale polynomially with the system size. 
Quantum simulation algorithms have been developed for in-medium jet evolution~\cite{Barata:2022wim,Barata:2023clv,Qian:2024gph,Barata:2026icn,Wu:2024adk,Castro:2025ocx} and the preparation of various hadron light-front wavefunctions~\cite{Kreshchuk:2020aiq,Kreshchuk:2020dla,Qian:2021jxp}.

Closely related to the present work are recent efforts to compute fragmentation functions, which encode the nonperturbative parton-to-hadron transition and cannot be easily obtained from first principles by conventional lattice methods~\cite{Metz:2016swz}, have been targeted in several reduced settings: as quasi-fragmentation functions from equal-time correlators in the massive Schwinger model~\cite{Grieninger:2024axp}, through a variational semi-inclusive hadron operator in the Nambu–Jona-Lasinio model~\cite{Li:2024nod}, and, most recently, directly in light-front QCD via real-time evolution, where the fragmentation was computed as a proof of concept~\cite{Galvez-Viruet:2025rmy}. These works all address vacuum fragmentation. Our study is complementary and distinct in two respects: first, we work with the (3+1)-dimensional light-front QCD Hamiltonian including transverse dynamics; second and most importantly, we compute hadron production in medium, with the quark-antiquark probe propagating through and scattering off the color field of a nuclear target before projection onto the hadron. This provides direct access to nuclear-medium observables, such as the dipole cross sections and the nuclear modification factor $R_A$, which are inaccessible to vacuum fragmentation calculations and are of direct relevance to the EIC.

In this work, we develop a quantum simulation framework for meson production in DIS using the light-front Hamiltonian formalism.
We simulate the real-time evolution of a $q\bar{q}$ dipole, both as a single-momentum state and as a component of the virtual photon, propagating through the color field of a nuclear target, and extract the meson production rate by projecting the evolved state onto hadron states.
The final hadron states are described by their light-front wavefunctions, and we focus here on the $J/\psi$ meson.
We consider the proton and the Au nuclear target, each characterized by their respective saturation scale, and extract the nuclear modification factor $R_A$ associated with the $J/\psi$ production rate.
 From the same framework, we also obtain the evolution of the dipole's momentum distribution, the total and elastic dipole cross sections, and the von Neumann entanglement entropy of the quark. 
Classical simulations are benchmarked against analytical eikonal predictions, and the full quantum-circuit implementation is demonstrated for the single-momentum dipole on a circuit emulator with a smaller lattice using \texttt{Qiskit}.

The paper is organized as follows: \cref{sec:th} introduces the theoretical setup including the light-front Hamiltonian, the initial states, the evolution scheme, and the key observables. \Cref{sec:qs} describes the quantum simulation strategy including qubit encoding, initial state preparation, Trotterized evolution, and quantum measurement protocols. \Cref{sec:results} presents the numerical results from both classical and quantum-circuit simulations.
We summarize and discuss future directions in \cref{sec:summary}.

\section{Theoretical methods\label{sec:th}}

\begin{figure*}[htp!]
    \centering
    \includegraphics[width=0.90\linewidth]{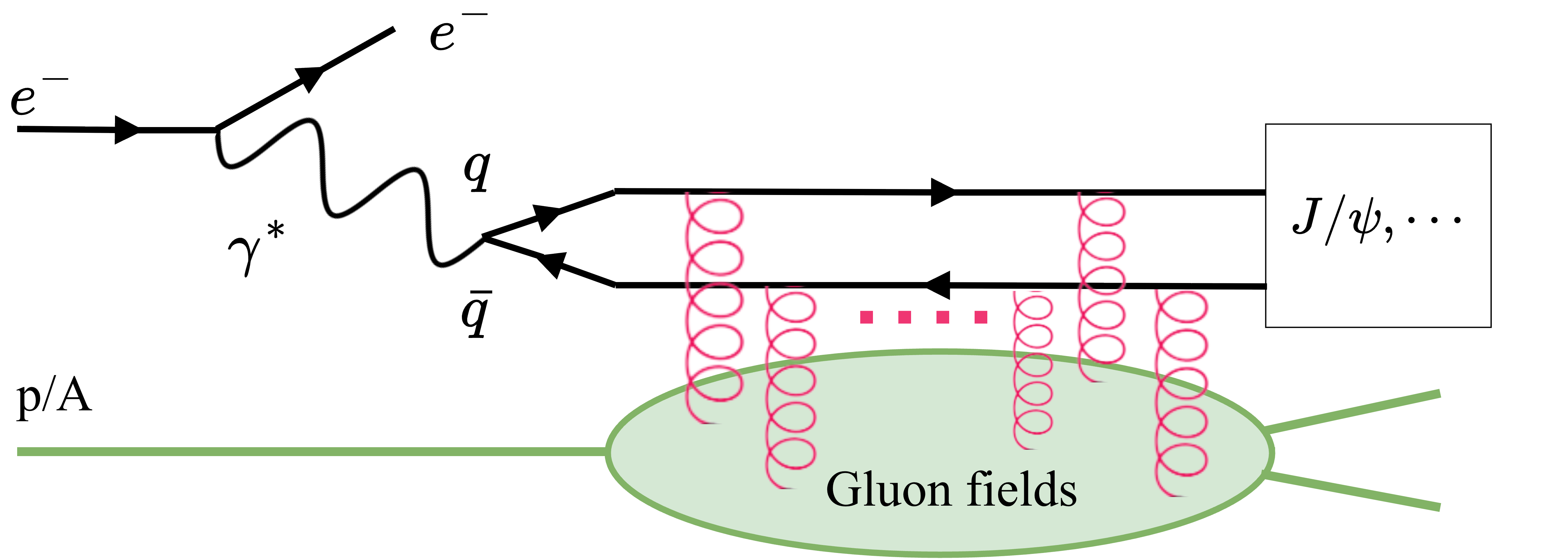}
    \caption{Diagram of the DIS physical process $\gamma^*+ p/A\rightarrow J/\psi+X$ where the proton $p$ or nucleus $A$ is taken as the background gluon field.}
    \label{fig:diagram}
\end{figure*}

The physical process of interest is represented in Fig.~\ref{fig:diagram}. Here, we study the deep inelastic scattering of an electron on a proton/nucleus target. Their exchanged virtual photon fluctuates into a quark-antiquark dipole, which further interacts with the field generated by the nucleus target. In the final stage, the dipole has the probability of combining into hadrons. We aim to calculate the dipole cross section, and the hadron production in such a process.

\subsection{The light-front QCD Hamiltonian}
In the physical process being considered, the dominant object is the dipole split from the virtual photon, and our focus is on studying the medium effect of the dipole. The Fock space expansion of the photon state can be written as
\begin{align}
   \ket{\gamma^*}_{phys} = \ket{\gamma^*}+\ket{q\bar q}+\ket{q\bar q g}+\ldots \;.
\end{align}
The light-front wavefunctions of the $q\bar q$ and $q\bar q g$ components, including quark masses, have been computed to the accuracy required for next-to-leading-order calculations in the dipole picture \cite{Beuf:2021qqa, Beuf:2021srj, Beuf:2022ndu}. As a first study, we restrict ourselves to the $\ket{q\bar q}  $ sector at leading order and leave the inclusion of higher sectors to future work.
As the dipole traverse the field generated by the nuclear target, its evolution is governed by the light-front QCD Hamiltonian containing the kinetic energy and background field interaction terms,
\begin{align}\label{eq:ham_evo}
\begin{split}
    P^-(x^+) &= P^-_\mathrm{KE} + V_\mathcal{A}(x^+)\\
    &=\left(P^-_{\mathrm{KE},q} + V_{\mathcal{A},q}(x^+)\right)
    +\left(P^-_{\mathrm{KE}, \bar{q}} + V_{\mathcal{A},\bar{q}}(x^+)\right)\;,
\end{split}
\end{align}
where $V_{\mathcal{A}}(x^+)$ denotes the interactions with the background field $ \mathcal A$. 

We choose a discrete momentum basis representation.
We consider a system contained within a box of finite volume \(\Omega = 2L \Omega_\perp\), with \(\Omega_\perp = (2L_\perp)^2\). We have introduced two spatial parameters: \(L\) in the longitudinal direction and \(L_\perp\) in the transverse directions. 
In the longitudinal direction, where \(-L \leq x^- \leq L\), we impose anti-periodic boundary conditions for fermions, which leads to a discretization,
  \begin{align}
      &p^+ = k^+ d_+, ~\text{with}~ k^+ = \frac{1}{2}, \frac{3}{2}, \ldots,\infty 
          \;,
  \end{align}
where the unit of $p^+$ is $d_+\equiv 2\pi/L$. The maximal value is restricted by the harmonic resolution $K=P^+ L/(2\pi)$, where $P^+$ is the total longitudinal momentum of the system.

In the transverse dimension,  $-L_\perp \le r^i\le L_\perp$, we impose the periodic boundary conditions and discretize the space into a square lattice,
  \begin{align}
    r^i= n_i a_\perp \quad (i = x,y)  \;,
 \end{align}
where $n_i=-N,-N+1,\ldots,N-1 $ and the lattice spacing is given by $a_\perp=L_\perp/N_\perp$. 
  The corresponding momentum space is also discrete with periodic boundary conditions,
  \begin{align}
     & p^i = k^i d_p \;,
  \end{align}
where $k^i = -N_\perp, -N_\perp+1 \ldots, N_\perp-1$ and $d_p\equiv \pi/L_\perp$ is the resolution in momentum space. 
    
The mode expansion of the field operators in such a discrete momentum basis is
    \begin{align}\label{eq:box_modes}
    \begin{split}
    \Psi_{\mu,c} (x)=  &\sum_{\bar{\alpha}} \frac{1}{\sqrt{ p^+ 2L (2L_\perp)^2}}
     [b^\pd_{\ind} u_\mu(p,\lambda)
       e^{-ip\cdot x}
       +d^\dagger_{\ind} v_\mu(p,\lambda) e^{ip\cdot x}]\;, 
    \end{split}
      \end{align}
    where $p\cdot x=1/2p^+ x^- -\vec p_\perp\cdot \vec x_\perp$ is the 3-product for the spatial components of $p^\mu$ and $x^\mu$.
  Each single particle state is specified by five quantum numbers, 
  \begin{align}
      \ind=\{ k^+, k^x, k^y, \lambda,c \}\;,
  \end{align}
   where $\lambda$ is the light-front helicity, and $c=1,2,\ldots, N_c$ is the color index of the quark/antiquark.
   For our problem in the $\ket{q\bar q}$ space, both the $p^+$ and the light-front helicity are preserved for each particle, so their values can be fixed throughout the evolution. Then the non-trivial quantum number are $\{ k^x, k^y, c \}$.
  The creation operators $b^{\dagger}_\ind$ and $d^{\dagger}_\ind$ 
  create quarks and antiquarks with quantum numbers $\ind$ respectively. 
  They obey the following anti-commutation relations,
  \begin{align}\label{eq:commutations}
    \{b^\pd_\ind,b^{\dagger}_{\ind'}\}=\{d^\pd_{\ind},d_{\ind'}^{\dagger}\}
=\delta^\pd_{\ind,\ind'} \;,  
  \end{align}
  with all the other anti-commutators vanishing.
The single-particle basis states are defined as 
  \begin{align}
  \begin{split}
       &\ket{q(\ind)}
    =b^\dagger_{\ind}\ket{0}\;,\\
           &\ket{\bar q(\ind)}
    =d^\dagger_{\ind}\ket{0}\;.
  \end{split}
  \end{align}
The multiple-particle basis states can be constructed as tensor products of the above states, e.g., the basis state of a dipole is $ \ket{q (\ind)\bar q(\ind')}=\ket{q(\ind)}\otimes \ket{\bar q(\ind')}$.

The dipole state is expanded in the basis space as
\begin{align}\label{eq:WF_coef}
    \ket{\psi_{q\bar q}}=\sum_\beta \sum_{\beta'} c_{\beta \beta'} \ket{q (\ind)\bar q(\ind')}\;,
\end{align}
where the basis coefficients $c_{\beta \beta'} $ consist into the dipole's light-front wavefunction (LFWF).
The state normalization reads
\begin{align}
  \sum_\beta \sum_{\beta'}|c_{\beta \beta'}|^2=1\;.
\end{align}

The second-quantized Hamiltonian operator in such basis can be written as~\cite{Qian:2024gph}
\begin{align}
\begin{split}
   P^-(x^+) = &\sum_{\ind} \frac{\vec{p}_{\perp}^2+m_q^2}{p^+} b_\ind^\dagger b^\pd_\ind + \sum_{\ind} \frac{\vec{p}_{\perp}^2+m_{ q}^2}{p^+}d_\ind^\dagger d^\pd_\ind \\
  &+ \frac{2g}{\Omega_\perp }\sum_{\ind_1,\ind_2} 
    \delta_{k_2^+,k_1^+}
    \delta_{\lambda_1,\lambda_2}
     T_{c_2,c_1}^a
     \tilde{\mathcal{A}}^a_+(\vec p_{2,\perp}- \vec p_{1,\perp}, x^+)
     b^\dagger_{\ind_2} 
    b^\pd_{\ind_1} 
     \\
  & -\frac{2g}{\Omega_\perp }\sum_{\ind_1,\ind_2} 
    \delta_{k_2^+,k_1^+}
    \delta_{\lambda_1,\lambda_2}
     T_{c_1,c_2}^a
     \tilde{\mathcal{A}}^a_+(\vec p_{2,\perp}- \vec p_{1,\perp}, x^+)
     d^\dagger_{\ind_2} 
    d^\pd_{\ind_1} 
   \;.
\end{split}
\end{align}
We adopt the McLerran-Venugopalan (MV) model~\cite{McLerran:1993ni, McLerran:1993ka, McLerran:1994vd} to describe the background field $\mathcal A$, and $\tilde{\mathcal{A}}$ is the field in the momemtum space via Fourier transform. The field is obtained by solving the reduced Yang-Mills equation
\begin{align}\label{eq:poisson}
 (m_g^2-\nabla^2_\perp )  \mathcal{A}^-_a(\vec x_\perp,x^+)=\rho_a(\vec x_\perp,x^+)\;,
\end{align}
where $m_g$ is the effective gluon mass of the field introduced to regulate the infrared divergence.
The color charge density source $\rho_a$ satisfies the local correlation function
\begin{align}\label{eq:MV_color_charge}
 \braket{\rho_a(\vec{x}_\perp,x^+)\rho_b(\vec{y}_\perp,y^+)}_{\text{event}}=g^2\tilde\mu^2\delta_{ab}\delta^2(\vec{x}_\perp-\vec{y}_\perp)\delta(x^+-y^+)\;.
\end{align}
Different samples of $\rho_a$ correspond to different configurations of the nuclear target, and we denote the average over these configurations by $\langle\cdots\rangle_\text{event}$. The background field is formulated in the same basis as the dipoles, and the correlator becomes~\cite{Lappi:2007ku,Li:2020uhl,Li:2021zaw}:
\begin{align}\label{eq:chgcor_dis}
    \braket{\rho_a(n_x,n_y,n_\tau)\rho_b({n'}_x,{n'}_y,n_\tau')}_{\text{event}}
    =g^2\tilde\mu^2
    \delta_{ab}\frac{\delta_{n_x,{n'}_x}\delta_{n_y,{n'}_y}}{a_\perp^2}\frac{\delta_{n_\tau,n_\tau'}}{\tau}\;,
\end{align}
where \(\vec x_\perp = (n_x, n_y) a_\perp\) and \(x^+ = n_\tau \tau \), with the layer index running over \( n_\tau = 1, 2, \ldots, N_\eta \). The total number of layers $N_\eta$ discretizes the evolution interval $x^+\in[0,L_\eta]$, so $\tau=L_\eta/N_\eta$. 
In practice, $\rho_a$ is sampled independently at each lattice site and layer from a Gaussian distribution with zero mean and variance $g^2\tilde\mu^2/(a_\perp^2\,\tau)$, and each sample yields one background-field configuration.

\subsection{Initial probe}\label{sec:initial_probe}

We consider two scenarios, a dipole with zero transverse momentum, and that produced from a virtual photon. The single-momentum dipole is the analytically-clean benchmark which can directly match with the Wilson-line eikonal formulas, while the virtual photon represents the physical probe in the dipole picture of DIS.

    \textbf{i. Single-momentum dipole}:
   We begin with a simplified setup where the dipole consists of a quark and an antiquark with definite momentum $ \vec p_{q,\perp}= \vec p_{\bar q,\perp}=\vec 0_\perp$, $p^+_q=p^+_{\bar q}=P^+/2$, in a color-singlet state.
   In terms of basis coefficients in \cref{eq:WF_coef}, the LFWF can be written as
   \begin{align}\label{eq:dipole_LFWF}
   c_{\beta\beta'}=\begin{cases}
        \dfrac{1}{\sqrt{N_c}}, & \text{for } k^i={\bar k^i}=0, z=1/2, k^+=\bar k^+=K/2, c=\bar c\\
        0, &\text{otherwise}
   \end{cases}
\end{align}
with $\beta=\{k^x, k^y,k^+, \lambda, c\}$ and $\beta'=\{\bar k^x, \bar k^y,\bar k^+, \bar \lambda, \bar c\}$ containing the quantum number of the quark and antiquark, respectively.

    \textbf{ii. Virtual photon}: The virtual photon wavefunction can be derived according to light-cone perturbation theory, to lowest order~\cite{Kowalski:2006hc,Dosch:1996ss,Lepage:1980fj,Li:2021cwv}, 
\begin{subequations}\label{eq:photon_LFWF}
\begin{align}\label{eq:photon_LFWF_mj0}
    \psi^{(m_j=0)}_{s\bar{s}/\gamma}(\vec k_\perp, x, Q)=  &\mathcal{Q}_f e\sqrt{N_c} \sqrt{x(1-x)}\frac{2 Q x(1-x)}{\epsilon^2+k_\perp^2}
    \delta_{s,-\bar{s}} ,\\
    \begin{split}
    \psi^{(m_j=1)}_{s\bar{s}/\gamma}(\vec k_\perp, x, Q) = &\mathcal{Q}_f e\sqrt{2N_c} \frac{\sqrt{ x(1-x)}}{\epsilon^2+k_\perp^2}
    \left[xk^R\delta_{s,+}\delta_{\bar{s}, -} - (1-x) k^R \delta_{s,-}\delta_{\bar{s}, +}\right.\\
    &
    \left. + m_f \delta_{s,+}\delta_{\bar{s}, +}\right],
    \end{split}\\
      \begin{split} \psi^{(m_j=-1)}_{s\bar{s}/\gamma}(\vec k_\perp, x, Q)= &-\mathcal{Q}_f e\sqrt{2N_c} \frac{\sqrt{ x(1-x)}}{\epsilon^2+k_\perp^2}
   \left [-xk^L\delta_{s,-}\delta_{\bar{s}, +} + (1-x) k^L \delta_{s,+}\delta_{\bar{s}, -} \right. \\
   &\left. +m_f \delta_{s,-}\delta_{\bar{s}, -}\right]\;,
    \end{split}
\end{align}
\end{subequations}
where $\mathcal{Q}_f e$ is the quark charge,  $Q^2$ the virtuality of the photon, and $\epsilon^2=x(1-x)Q^2 + m_f^2$ the energy. The relative momentum is written as $k^R=k^x+ik^y$ and $k^L=k^x-ik^y$. Specifically, $\mathcal{Q}_c = + 2/3$ for the charm quark and $\mathcal{Q}_b=-1/3$ for the bottom quark. 
Since the photon is color neutral, the color part of the dipole LFWF is a color-singlet state, $    (1/\sqrt{N_c})\sum_{c=1}^{N_c}\sum_{\bar c=1}^{N_c}\delta_{c,\bar c} $,
which is common to all three polarization components and has been suppressed in \cref{eq:photon_LFWF} for brevity.
The photon LFWF can be written equivalently in the single-particle basis with 
\begin{align}\label{eq:rel_2_single}
    \begin{split}
       & p^+_q=xP^+\;,
        \qquad  p^+_{\bar q}=(1-x)P^+\;,\\
       & \vec p_{\perp,q} = x\vec P_\perp + \vec k_\perp\;,\qquad 
       \vec p_{\perp,\bar q} = (1-x)\vec P_\perp - \vec k_\perp\;.
    \end{split}
\end{align}
The reverse coordinate transformation reads,
\begin{align}\label{eq:CMP_k}
    \begin{split}
       & P^+=p^+_q +  p^+_{\bar q}\;,\\
       & \vec P_\perp =\vec p_{\perp,q} +  \vec p_{\perp, \bar q}\;,\qquad \vec k_\perp=(1-x)\vec p_{\perp,q} -x\vec p_{\perp,\bar q} \;.
    \end{split}
\end{align}
Substituting these into \cref{eq:photon_LFWF} gives the photon LFWF in the single-particle representation,
\begin{align}
\tilde  \psi^{(m_j)}_{s\bar{s}/\gamma}(\vec p_{q,\perp}, p^+_q, \vec p_{\bar q,\perp}, p^+_{\bar q},Q)=
 \psi^{(m_j)}_{s\bar{s}/\gamma}(\vec k_\perp, x, Q)\;,
\end{align}
where $x$ and $\vec k_\perp$ on the right-hand side are functions of the
single-particle momenta through \cref{eq:CMP_k}. The change of variables only
relabels the arguments of the wavefunction and introduces no additional
factor.

In the discrete momentum basis, the transverse momenta live on a square lattice with periodic boundary conditions, which complicates the transformation in two ways. First, the relative momentum in units of the
lattice resolution is in general not an integer vector, so that $\vec k_\perp$ does not fall on the lattice sites. Second, the total transverse momentum is conserved only up to a multiple of the lattice period, so that $\vec k_\perp$ is not uniquely fixed by \cref{eq:CMP_k}. We treat both issues following the prescription detailed in Appendix~D of \cite{Li:2025wzq}.\footnote{ 
The prescription is that of Appendix~D of \cite{Li:2025wzq}, with the gluon and quark quanta therein replaced by quark and antiquark quanta in the present case.
Concretely, on the discrete lattice with periodic boundary (P.B.) conditions, the momentum-conserving delta is promoted to
\begin{align}
\delta_{\vec P_{\perp}, \vec p_{q,\perp} + \vec p_{\bar q,\perp} }\to \delta_{\vec P_{\perp}, P.B.[\vec p_{q,\perp} + \vec p_{\bar q,\perp}] }\;,
\end{align}
where $\mathrm{P.B.}[\vec v\,]$ denotes the representative of $\vec v$ folded back into the fundamental period of the lattice.
}
The longitudinal momentum is discretized into $K$ portions, and the accessible values of $x$ are $x=k^+_q/K$ with $k_q^+=1/2,3/2,\cdots,(2K-1)/2$.

For photon state with a fixed polarization, its LFWF can be written in terms of basis coefficient 
\begin{align}
    \ket{\psi^{(m_j)}_{\gamma}(P)}= \sum_\beta \sum_{\beta'}
    c^{\gamma,(m_j)}_{\beta \beta'}\ket{q (\beta)\bar q(\beta')}\;.
\end{align}
Note that the photon state here is not normalized to unity, since only its $q\bar q$ component, obtained from perturbative expansion, is included.
The continuum momentum state is related to the discrete momentum basis state as
\begin{align}
    \ket{\vec p_\perp, p^+}=\sqrt{p^+ \Omega}\ket{\beta(\vec k_\beta=\vec p_\perp/d_p, k^+_\beta=p^+/d_+)}\;.
\end{align}
It follows that, 
\begin{align}
\begin{split}
   c^{\gamma,(m_j)}_{\beta \beta'}
   =&\frac{1}{\sqrt{ \Omega P^+ x(1-x)}}
   \tilde  \psi^{(m_j)}_{s\bar{s}/\gamma}(\vec p_{q,\perp}, p^+_q, \vec p_{\bar q,\perp}, p^+_{\bar q}) \delta_{p_q+p_{\bar q},P}\\
   =&\frac{1}{\sqrt{4\pi K (2L_\perp)^2 x(1-x)}}
   \tilde  \psi^{(m_j)}_{s\bar{s}/\gamma}(\vec p_{q,\perp}, p^+_q, \vec p_{\bar q,\perp}, p^+_{\bar q}) \delta_{p_q+p_{\bar q},P}
   \;.
\end{split}
\end{align}

\subsection{Unitary evolution}

The evolution of the dipole is governed by the time evolution equation on the light front,
\begin{align}
  \label{eq:ShrodingerEq}
  i\frac{\partial}{\partial x^+}\ket{\psi;x^+}=\frac{1}{2}P^-(x^+)\ket{\psi;x^+}\;.
\end{align}
 The solution of~\cref{eq:ShrodingerEq} describes the state of the investigated system at any later light-front time $x^+$,
\begin{align}\label{eq:ShrodingerEqSol}
  \ket{\psi;x^+}=\mathcal{T}_+\exp\left[-\frac{i}{2}\int_0^{x^+}\diff z^+P^-(z^+)\right]\ket{\psi;0}\;,
\end{align}
where $\mathcal{T}_+$ is the light-front time ordering.

A common approach to non-perturbative simulation of time evolution in~\cref{eq:ShrodingerEqSol} amounts to trotterize the light-front time into a large number of small steps as
\begin{align}\label{eq:time_evolution_exp}
 \begin{split}
   \mathcal{T}_+ &\exp[-\frac{i}{2}\int_0^{x^+}\diff z^+P^-(z^+)]
   =\lim_{n\to\infty}\prod^{n-1}_{k=0}\mathcal{T}_+ \exp\left[-\frac{i}{2}\int_{x_{k}^+}^{x_{k+1}^+}\diff z^+P^-(z^+)\right]
 \;,
 \end{split}
\end{align}
where the step size is $\delta x^+ \equiv x^+/n$, and the intermediate times are $x_k^+=k\delta x^+ (k=0,1,2,\ldots,n)$ with $x_0^+=0$ and $x_n^+=x^+$. 
Within each short timestep, the Hamiltonian can be considered time-independent, such that the evolution operator $U_k=U(x_k^+;x_{k+1}^+)$ is written as 
\begin{align}
 \begin{split}\label{eq:trotter_1st}
   U_k\equiv &\lim_{\delta x^+\to 0}\mathcal{T}_+ \exp\left[-\frac{i}{2}\int_{x_{k}^+}^{x_{k+1}^+}\diff z^+P^-(z^+)\right]
  =\exp\left[-\frac{i}{2} P^-(x_k^+) \delta x^+\right]\;.
 \end{split}
\end{align}
In classical simulations, the dipole state and the evolution operator are represented as a column vector and a matrix, correspondingly.

The evaluation of physical observables or quantities is straightforward by applying the corresponding operator to the state vector. For instance, given an operator $\hat O$ and the quantum state $\ket{\psi}$, the expectation value can be expressed as: 
\begin{align}\label{eq:expectation_measure}
    \braket{\hat O}_{\psi} = \braket{\psi|\hat O|\psi}\;.
\end{align}
To account for the fluctuation introduced by the background field, which physically corresponds to the configuration fluctuation of the nuclear target in different experimental events, the obtained quantities should be averaged over multiple medium configurations,
\begin{align}
  \braket{\hat O} = \braket{ \braket{\hat O}_{\psi}}_{\text{event}}\;.
\end{align}

\subsection{Dipole cross sections}

The cross section is defined as 
\begin{align}\label{eq:cross}
    \begin{split}
    \frac{\diff\sigma}{\diff^2 b} 
    =&\sum_{\phi_f} |M(\phi_f;\psi_{in})|^2\;,
\end{split}
\end{align}
in which $\sum_{\phi_f}$ is the summation over the final phase space. The transition amplitude can be written as
\begin{align}\label{eq:M}
    \begin{split}
M(\phi_f;\psi_{in})
    =&\braket{\phi_f|S|\psi_{in}}-\braket{\phi_f|\psi_{in}}\\
    =&\braket{\phi_f|\psi_{out}}-\braket{\phi_f|\psi_{in}}\;.
\end{split}
\end{align}

The $S$ in the equation is the evolution operator from the initial state to the final through a finite time transition, and is different from the case where one takes the infinite time limit.
The incoming state is the initial state, and the outgoing state is the final state after the evolution of duration $L_\eta$,
\begin{align}
    \ket{\psi_{in}}=   \ket{\psi(x^+=0)}\;,\qquad
     \ket{\psi_{out}}=   \ket{\psi(x^+=L_\eta)}\;.
\end{align}

The evaluation of the total and the elastic cross section differs in the order of event average and amplitude square~\cite{Mueller:1997ik,Kovchegov:1999kx,Dumitru:2002qt},
\begin{subequations}\label{eq:crss_defs}
\begin{align}
    &\frac{\diff\sigma_{\mathrm{tot}}}{\diff^2 b}
    =\langle\sum_{\phi_f}{|  M(\phi_f; \psi_{in})|}^2\rangle_{\text{event}}\;,\\
       & \frac{\diff\sigma_{\mathrm{el}}}{\diff^2 b}
    =\sum_{\phi_f}{|  \langle M(\phi_f; \psi_{in}) \rangle_{\text{event}}|}^2
    \;.
\end{align}
\end{subequations}
The total cross section is obtained by averaging the squared scattering amplitude over target configurations, whereas the elastic cross section is obtained by squaring the event-averaged amplitude.

In the eikonal limit $p^+=\infty$, the cross sections of a color-singlet $q\bar q$ dipole with quark at $\vec{x}_\perp$ and antiquark at $\vec{y}_\perp$ can be evaluated analytically in the MV model~\cite{McLerran:1993ni,McLerran:1993ka} (see also Ref.~\cite{Kovchegov:2012mbw, Fukushima:2007dy} for a pedagogical derivation),
\begin{subequations}
\label{eq:crss_eikonal_qqbar_full}
\begin{align}
  \begin{split} \label{eq:crss_eikonal_qqbar_total}
   \frac{\diff\sigma_{q\bar q,\mathrm{tot}}}{\diff^2 b}\bigg|_{p^+=\infty}
     =&2\bigg\{
      1-
      \exp\bigg[
      - C_F
      \frac{g^4\tilde\mu^2 L_\eta}{4\pi m_g^2}
      \left[
        1-
        m_g|\vec x_\perp-\vec y_\perp| K_1(m_g|\vec x_\perp-\vec y_\perp|)
      \right]
      \bigg]
    \bigg\}
    \;,
  \end{split}\\
  \label{eq:crss_eikonal_qqbar_elastic}
     \frac{\diff\sigma_{q\bar q,\mathrm{el}}}{\diff^2 b}\bigg|_{p^+=\infty}
     =&\bigg\{
      1-
      \exp\bigg[
      - C_F
      \frac{g^4\tilde\mu^2 L_\eta}{4\pi m_g^2}
      \left[
        1-
        m_g|\vec x_\perp-\vec y_\perp| K_1(m_g|\vec x_\perp-\vec y_\perp|)
      \right]
      \bigg]
    \bigg\}^2\;,
\end{align}
\end{subequations}
where $C_F=(N_c^2-1)/(2N_c)$, $K_1$ is the modified Bessel function, and $g^2\tilde\mu$ and $m_g$ are the charge density and IR regulator of the MV model defined in \cref{eq:MV_color_charge,eq:poisson}.
The correlation between the quark and the antiquark is stronger when the separation between them $|\vec x_\perp - \vec y_\perp|$ is smaller. 
In the coincidence limit $|\vec x_\perp - \vec y_\perp|=0$, the cross section is zero.
In the limit $|\vec x_\perp - \vec y_\perp|=r\to\infty$, $ r K_1(r)\to 0$, the cross section becomes
\begin{align}\label{eq:crss_rinf}
  \begin{split}
   \frac{\diff\sigma_{q\bar q,\mathrm{tot}}}{\diff^2 b}\bigg|_{p^+=\infty, |\vec x_\perp - \vec y_\perp|\to\infty }
     =&2\bigg\{
      1-
      \exp\bigg[
      - C_F
      \frac{g^4\tilde\mu^2 L_\eta}{4\pi m_g^2}
      \bigg]
    \bigg\}
    \;.
  \end{split}
\end{align}
For a dipole state that is localized in momentum space, i.e., $ \vec p_{q,\perp}=\vec p_{\bar q,\perp}=\vec 0_\perp$, their separation in the position space is large, and the cross section can be approximated by Eq.~\eqref{eq:crss_rinf}. For a more accurate estimation, one can use
   \begin{align}\label{eq:dipole_r_estimate}
  \frac{1}{(2L_\perp)^4}\int \diff^2 x_\perp \diff^2 y_\perp
   \frac{\diff\sigma_{q\bar q,\mathrm{tot}}(|\vec x_\perp - \vec y_\perp|)}{\diff^2 b}\bigg|_{p^+=\infty}\;.
\end{align}   
The integration over the transverse space can be replaced by summation over the transverse lattice sites. 
\subsection{Hadron production}\label{sec:hadron_production_method}
With the evolved state $\psi(x^+)$, the probability to produce a hadron $h$ is obtained by projecting it onto the hadron state.
As the dipole can also acquire different C.M. momenta due to the momentum transfer from the nuclear target, the projection is carried out for each possible final momentum of the hadron, and the resulting probabilities are then summed over. 
We can therefore write the probability as,
\begin{align}\label{eq:meson_prob}
    \mathrm{Prob}(h) = \sum_{\vec P_\perp}|\braket{\psi(x^+) | \psi_h(\vec P_\perp)}|^2\;,
\end{align}
where $\ket{\psi_h(\vec P_\perp)}$ is the hadron state with transverse C.M. momentum $\vec P_\perp$.

We adopt the charmonium LFWF proposed in Ref.~\cite{Li:2021cwv}.
The $J/\psi$ LFWF reads
\begin{subequations}
  \label{eq:LFWF_Jpsi}
  \begin{align}
    &\psi^{(m_j=0)}_{J/\psi} = 
   \psi_{0,0,0} \sigma_+
    \;,
    \label{eq:WF_base1_mj0}\\
    &
    \psi^{(m_j=1)}_{J/\psi} = 
   \psi_{0,0,0}\sigma_{\uparrow\uparrow}
    \;,
    \label{eq:WF_base1_mj1}\\
    &
    \psi^{(m_j=-1)}_{J/\psi} = 
    \psi_{0,0,0}\sigma_{\downarrow\downarrow}
    \;.
    \label{eq:WF_base1_mjm1}
  \end{align}
\end{subequations}
We define $\sigma_{\pm}\equiv 1/\sqrt{2}(\sigma_{\uparrow\downarrow}\pm\sigma_{\downarrow\uparrow})$ for convenience. The $nml$-basis function is 
\begin{align}\label{eq:psi_nml}
  \psi_{n,m,l}(\vec k_\perp,x)\equiv  \phi_{nm}(\vec k_\perp/\sqrt{x(1-x)}) \chi_l(x)
  \;.
\end{align}
The transverse basis function in the momentum space reads
\begin{align}\label{eq:phinm}
  \begin{split}
  \phi_{nm}(\vec k_\perp)=&\kappa^{-1}\sqrt{\frac{4\pi n!}{(n+|m|)!}}{\bigg(\frac{k_\perp}{\kappa}\bigg)}^{|m|}
  \exp(-k_\perp^2/(2\kappa^2))L_n^{|m|}({k_\perp}^2/\kappa^2)\exp(im\theta_k)
  \;,    
  \end{split}
\end{align}
where $k_\perp=|\vec k_\perp |$, $\theta_k=\arg (k^x+ik^y)$, $n=0,1,2,\ldots$ is the principal number, and $m=0,\pm1,\pm2,\ldots$ is the orbital number. 
The transverse confining strength $\kappa=1.34 \GeV$ functions as a basis parameter here.
The longitudinal basis function is 
\begin{align}
  \begin{split}
  \chi_l(x)=&\sqrt{4\pi(2l+\alpha+\beta+1)}\sqrt{\frac{\Gamma(l+1)\Gamma(l+\alpha+\beta+1)}{\Gamma(l+\alpha+1)\Gamma(l+\beta+1)}}
  x^{\beta/2}{(1-x)}^{\alpha/2}P_l^{(\alpha,\beta)}(2x-1)
  \; ,    
  \end{split}
\end{align}
where $P_l^{(\alpha,\beta)}$ is the Jacobi polynomial with $l=0,1,2,\ldots$. 
The dimensionless basis parameters $\alpha$ and $\beta$ are related to the longitudinal confining strength and the fermion mass as $\alpha =4m_q^2/\kappa^2$ and $\beta = 4m_{\bar q}^2/\kappa^2$. For charmonia, $m_q=m_{\bar q}=m_f=1.27 \GeV$.

We map the meson LFWF onto the discrete basis representation following the same procedure as for the photon.
The basis coefficients are obtained from the continuum wave function in \cref{eq:LFWF_Jpsi}, and the resulting state is then
normalized,
\begin{align}
    \ket{\psi^{(m_j)}_{J/\psi}(\vec P_\perp)}=\mathcal N \sum_\beta \sum_{\beta'} 
    c^{J/\psi,(m_j)}_{\beta \beta'}\ket{q (\ind)\bar q(\ind')}\;.
\end{align}
As in the photon case, the basis coefficients are related to the continuum wavefunction by
\begin{align}\label{eq:Jpsi_coef}
\begin{split}
c^{J/\psi,(m_j)}_{\beta\beta'}
&=\frac{1}{\sqrt{\Omega P^+x(1-x)}}\,
\tilde\psi^{(m_j)}_{s\bar s/J/\psi}(\vec p_{q,\perp},p^+_q,\vec p_{\bar q,\perp},p^+_{\bar q})\,
\delta_{p_q+p_{\bar q},P}\\
&=\frac{1}{\sqrt{4\pi K(2L_\perp)^2x(1-x)}}\,
\psi^{(m_j)}_{s\bar s/J/\psi}(\vec k_\perp,x)\,
\delta_{p_q+p_{\bar q},P}\;,
\end{split}
\end{align}
where the relative momentum $(\vec k_\perp, x)$ is determined by the single-particle momenta carried by the basis states $\beta$ and $\beta'$ through \cref{eq:rel_2_single}. Since the continuum LFWF is normalized to unity, $\mathcal N$ differs from one only by discretization effects.

To study the nuclear modification effect, we define the nuclear modification factor of the meson production rate,
\begin{align}
    R_{A}=
    \frac{\mathrm{Prob}_{e+A}(J/\psi) }{\mathrm{Prob}_{e+p}(J/\psi)}\;.
\end{align}
The charge density of the nuclei is fixed, and the size of the nuclei, which is also the duration of the background field along $x^+$, scales according to nuclear geometry, 
\begin{align}
    L_{\eta,A}=A^{1/3}L_{\eta,p}\;,
\end{align}
with $L_{\eta,p}\sim 1.2 $ fm$=6 \GeV^{-1}$.
Since the transverse interaction area is the same for both targets, with only the longitudinal extent varied according to nuclear geometry, $R_A$ as defined here is already a per-unit-area ratio and no additional normalization is implemented.
We consider EIC kinematics for proton and Au-197 nucleus target, so $L_{\eta,Au}\sim 7 $ fm$=35 \GeV^{-1}$.
The saturation scale is defined as $Q_{s}^2 =C_F (g^2\tilde\mu)^2 L_\eta/(2\pi^2)$.
We take $g^2\tilde\mu=0.8\GeV^{3/2}$, such that $Q_{s,p}=0.51\GeV$ for proton target, then $Q_{s,Au}=1.23 \GeV$ for Au target.

\section{Quantum simulation strategies}\label{sec:qs}

In this section, we discuss the specific procedure to simulate the real-time evolution of the quark-antiquark probe in medium on a digital quantum computer and the protocol to extract various physical observables as mentioned in the previous section.
\begin{figure*}[h]
    \centering
    \includegraphics[width=0.70\linewidth]{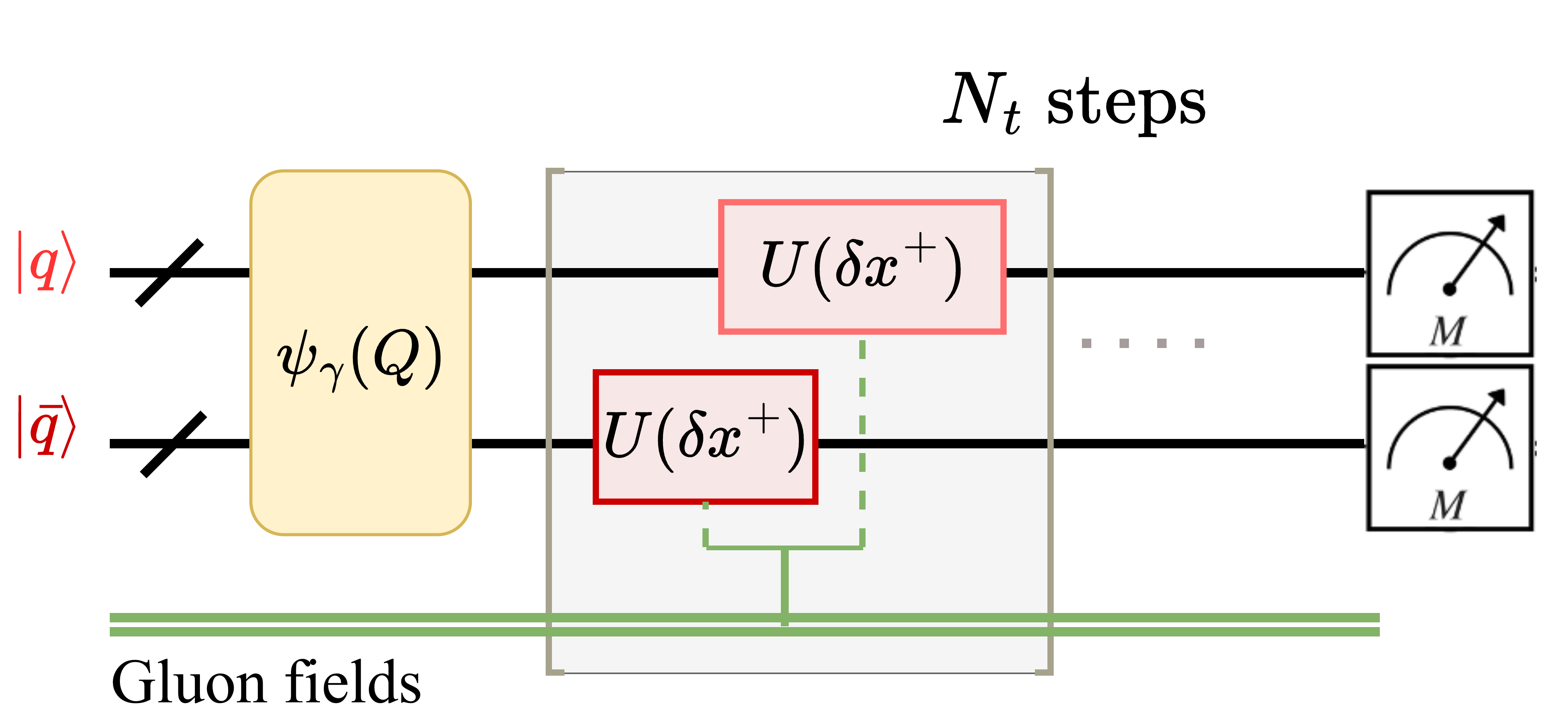}
    \caption{Schematic of the quantum circuit for the real-time probe evolution in background medium in the $\ket{q\bar{q}}$ Fock space.}
    \label{fig:diagram_qc}
\end{figure*}

The simulation process contains several key components including 1) encoding the Hamiltonian and quantum state onto qubits, 2) initial state preparation, 3) simulating real-time dynamics using evolution operator, and 4) measurement and extracting information from final quantum state. In Fig.~\ref{fig:diagram_qc}, we provide an illustration for the full simulation framework and explain each step in more details in the following subsections.

\subsection{Qubit encoding of the Hamiltonian and basis state}\label{subsec:encoding}

First and foremost, we describe the procedure to encode the light-front QCD Hamiltonian and its basis states onto qubits following the prescriptions introduced in earlier works~\cite{Qian:2024gph, Barata:2026icn}. Other related quantum simulation using light-front Hamiltonian approaches are also found in Refs.~\cite{Barata:2021yri,Barata:2022wim,Barata:2023clv,Wu:2024adk, Castro:2025ocx,Galvez-Viruet:2024hry}. In this work, we use the direct encoding scheme based on the second-quantized Hamiltonian where individual degree of freedom is mapped directly onto respective qubit. This approach proves to be efficient in constructing the necessary quantum circuit and scales polynomially with system size.

In the direct encoding scheme, we map the finite-dimensional fermionic Fock space to a multi-qubit system. In this way, each basis mode $\beta_n$ in physical Hamiltonian is uniquely mapped to some qubit index $n$. Therefore the total number of qubits for the simulation is $N_q=2N_F$ where the $N_F$ is the number of fermion and anti-fermion modes. Within the $\ket{q\bar{q}}$ Fock sector, the longitudinal quantum number is not modified throughout the evolution, therefore we can assign different $x$ values for the initial probe to save the number of qubits. The same reason is also applied to spin component. Since our momentum space is discretized from $-N_\perp$ to $N_\perp-1$ for both transverse directions, the total number of qubits $N_q = 2N_F = 2 N_c(2N_\perp)^2=8N_c N_\perp^2$ for the main circuit of the evolution. Specifically, we choose to map the fermion and antifermion degrees of freedom onto distinct block of the qubits such that the $\ket{q\bar{q}}$ Fock sector iterates over all the possible $\ket{q}\otimes \ket{\bar{q}}$ on the qubits. 

To satisfy the anti-commutativity of fermionic creation and annihilation operators, one can use the standard Jordan-Wigner (JW) fermion-to-qubit mappings~\cite{Jordan:1928wi},
\begin{align}
\label{eq:JW}
&b_n^\dagger \mapsto  \frac{X_n-iY_n}{2}\;,\quad b_n \mapsto  \frac{X_n+iY_n}{2}\;, \\
&d_n^\dagger \mapsto  \frac{X_n-iY_n}{2}\;,\quad d_n \mapsto  \frac{X_n+iY_n}{2}\;, 
\end{align}
where $X_n, Y_n, Z_n$ are the Pauli-X, Pauli-Y, and Pauli-Y matrix on qubit $n$. 
Note that we have excluded the Pauli-Z matrices that are relevant to exchange of fermions and antifermions because our Hamiltonian $P^-(x^+)$ strictly conserves $\ket{q\bar{q}}$ Fock sector. Since our Hamiltonian consists of only $b^\dagger_n b_m$ and $d^\dagger_n d_m$ terms, we only have two-qubit Pauli-matrix terms and their total number scales with $\mathcal{O}(N_q^2)$, which makes it efficient in simulating the unitary evolution operator on the circuit.

\subsection{Preparation for initial probe}

In this work, we consider two initial states for the computation, as formulated in \cref{sec:initial_probe}: one is the single-momentum dipole, the other the photon state. The preparation of the photonic initial state in $\ket{q\bar{q}}$ Fock sector can be challenging, which may involve the construction of specific unitary ansatzes or quantum adiabatic state preparation techniques. Here, we focus on the preparation of the single-momentum dipole.
\begin{figure}[t]
\centering
\[
\begin{quantikz}
  \lstick{$q_1  \colon \ket{0}$} & \gate{X} & \qw      & \targ{}   & \qw       & \qw        & \qw \\
  \lstick{$q_2 \colon \ket{0}$} & \qw      & \gate{H} & \ctrl{-1} & \ctrl{1}  & \ctrl{2}   & \qw \\
  \lstick{$\bar{q}_1  \colon \ket{0}$} & \gate{X} & \qw      & \qw       & \targ{}   & \qw        & \qw \\
  \lstick{$\bar{q}_2 \colon \ket{0}$} & \qw      & \qw      & \qw       & \qw       & \targ{}    & \qw
\end{quantikz}
\quad
\begin{array}{l}
  \left.\rule{0pt}{5.5ex}\right\} \text{$q$ registers} \\[5.5ex]
  \left.\rule{0pt}{5.5ex}\right\} \text{$\bar{q}$ registers}
\end{array}
\]
\caption{Quantum circuit preparing the singlet state $\ket{\psi_\mathrm{singlet}}$ in SU(2). }
\label{fig:stateprep_singlet}
\end{figure}
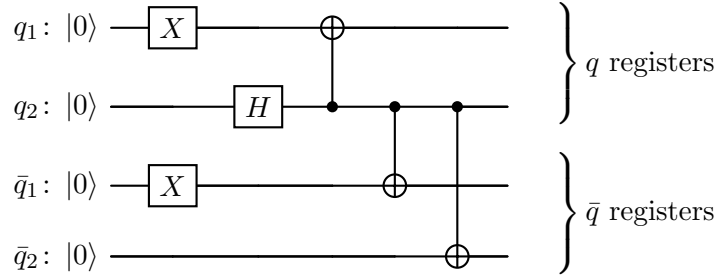
The singlet $\bm{p}=0$ momentum state in $\ket{q\bar{q}}$ is
\begin{align}\label{eq:psi_s}
    \ket{\psi_\mathrm{singlet}(\bm{p}=\bm{0})} = U_{q\bar{q}} \ket{0}^{\otimes N_q} =  \frac{1}{\sqrt{N_c}}\sum_{c=1}^{N_c}\ket{q(\bm{0}, c) \bar q(\bm{0}, c)}\;,
\end{align}
and its preparation using direct encoding can be implemented straightforwardly on the quantum circuit. For SU(2), one simplest possible unitary circuit $U_{q\bar{q}}$ for the initial state can be constructed using an Hadamard gate, two X-gates, and three CX-gates, as illustrated in \cref{fig:stateprep_singlet}. 
Here, the qubits $q_1$ and $q_2$ represent the quark mode $q(\bm{0},c=0)$ and $q(\bm{0},c=1)$,  respectively, while $\bar{q}_1$ and $\bar{q}_2$ represent the corresponding antiquark modes $\bar{q}(\bm{0},c=0)$ and $\bar{q}(\bm{0},c=1)$.

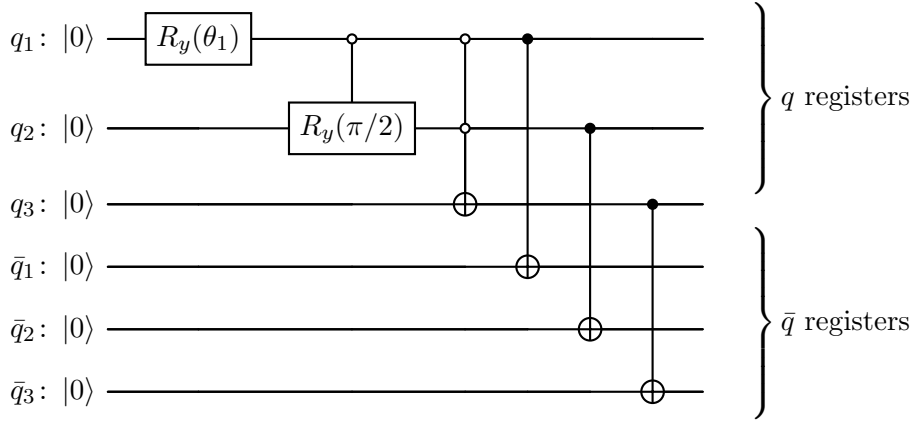
\begin{figure}[t]
\centering
\[
\begin{quantikz}
  \lstick{$q_1   \colon \ket{0}$}     & \gate{R_y(\theta_1)} & \octrl{1}         & \octrl{2} & \ctrl{3} & \qw      & \qw      & \qw \\
  \lstick{$q_2   \colon \ket{0}$}     & \qw                  & \gate{R_y(\pi/2)} & \octrl{1} & \qw      & \ctrl{3} & \qw      & \qw \\
  \lstick{$q_3   \colon \ket{0}$}     & \qw                  & \qw               & \targ{}   & \qw      & \qw      & \ctrl{3} & \qw \\
  \lstick{$\bar{q}_1 \colon \ket{0}$} & \qw                  & \qw               & \qw       & \targ{}  & \qw      & \qw      & \qw \\
  \lstick{$\bar{q}_2 \colon \ket{0}$} & \qw                  & \qw               & \qw       & \qw      & \targ{}  & \qw      & \qw \\
  \lstick{$\bar{q}_3 \colon \ket{0}$} & \qw                  & \qw               & \qw       & \qw      & \qw      & \targ{}  & \qw
\end{quantikz}
\quad
\begin{array}{l}
  \left.\rule{0pt}{8.5ex}\right\} \text{$q$ registers} \\[8.5ex]
  \left.\rule{0pt}{8.5ex}\right\} \text{$\bar{q}$ registers}
\end{array}
\]
\caption{Quantum circuit preparing the singlet state $\ket{\psi_\mathrm{singlet}}$ in SU(3). The rotation angle is $\theta_1 = 2\arcsin(1/\sqrt{3})$. Open (filled) circles denote $\ket{0}$-controls (anti-controls) and $\ket{1}$-controls, respectively.
Columns 1-3 prepare the W state $\ket{W_3}$ on the quark register; columns 4-6 entangles the quark to the antiquark sector via CNOT gates.
}
\label{fig:stateprep_singlet_su3}
\end{figure}

In general, for $\mathrm{SU}(N_c)$ with $N_c\geq 3$, the singlet state can also be systematically constructed with the help of W states~\cite{Barenco:1995na,Cruz:2018oka,Dur:2000zz}.
Starting from $\ket{0}^{\otimes N_c}$ on the quark register, and one sequentially applies the controlled-rotation $R_y(\theta_k)$ with $k = 1, \ldots, N_c-1$ on qubit $q_k$, conditioned on all preceding qubits being in $\ket{0}$, with angles
\begin{align}\label{eq:wstate_angles}
  \theta_k = 2\arcsin\!\left(\frac{1}{\sqrt{N_c - k + 1}}\right), \qquad k = 1, \ldots, N_c - 1\,,
\end{align}
followed by a $(N_c-1)$-fold anti-controlled CNOT on $q_{N_c}$.
The angles in Eq.~\eqref{eq:wstate_angles} are chosen such that $\sin^2(\theta_k/2) = 1/(N_c - k+1)$,
which at each step $k$ redistributes the remaining amplitude $\sqrt{(N_c-k)/N_c}$ uniformly over the $N_c - k$ unset color modes. After the W state $\ket{W_{N_c}}$ is established on the quark register, $N_c$ CNOT gates entangles the color occupation to the antiquark sector, completing the singlet preparation. As an example, we show the quantum circuit to prepare the singlet in \cref{fig:stateprep_singlet_su3} with a single Toffoli gate.

\subsection{Real-time unitary evolution}
Now that the initial state of the probe is prepared, one can simulate its time-dependent Hamiltonian evolution. 
The evolution is implemented on the quantum circuit through a Trotterized unitary evolution, in the same scheme as written in \cref{eq:time_evolution_exp},
\begin{align}\label{eq:uni_evo}
\begin{split}
    \ket{\psi(x^+)} &= U(x^+, 0) \ket{\psi(0)}
  = \cT_+ e^{-\frac{i}{2}\int_{0}^{x^+} \diff z^+\, P^-(z^+) } \ket{\psi(0)} \\
    &= \prod_{k=0}^{N_t-1} U(x_{k+1}^+, x_{k}^+) \ket{\psi(0)}\;,
\end{split}
\end{align}
where $x_0^+=0, x_{N_t}^+=L, \delta x^+ = x_{k+1}^+ - x_{k}^+= L/N_t$, and $N_t$ is the number of Trotter steps. 
For sufficiently small $\delta x^+$, the Hamiltonian can
be regarded as constant within each time step, yielding
\begin{align}\label{eq:uni_evo_step}
\begin{split}
   U(x_{k+1}^+, x_{k}^+) 
\approx
e^{-\frac{i}{2} P^-(x_{k}^+) \delta x^+}\;.
\end{split}
\end{align}
Within each trotter time step, one needs to consider $\mathcal{O}(N_q^2)$ two-qubit Pauli matrices, leading to $\mathcal{O}(N_q^2)$ CNOT gates per step. Using a $p$th-order Trotter formula, the overall gate complexity scales asymptotically as $\mathcal{O}( N_q^2 t^{1+1/p}/\epsilon_{\text{trotter}}^{1/p})$ for an evolution time $t$ and trotter error $\epsilon_{\text{trotter}}$ for the entire process. 

Alternatively, one can use quantum split-operator Fourier transform (Q-SOFT)~\cite{zalka1998simulating, kassal2008polynomial, cao2019quantum} to simulate the time evolution Hamiltonian by separating the kinetic energy part and the background interaction part using a quantum Fourier transform (qFT) that transforms between momentum basis and position basis~\cite{Barata:2023clv}. This is advantageous because the kinetic energy operators are diagonal in momentum basis and the medium interaction operators are diagonal in position basis, reducing the total number of Pauli matrices to $\mathcal{O}(N_q)$. The fermonionic Fourier transforms require an additional $\mathcal{O}(N_q \log N_q)$ gates~\cite{babbush2018low,kivlichan2018quantum} resulting in a better total cost of $\mathcal{O}(N_q\log N_q)$ for the gate counts. Nonetheless, in this work, we choose to simulate the time evolution in the momentum space without the qFT for the small number of qubits less than $\mathcal{O}(100)$ considered in our setup.

\subsection{Measurement and extraction of observables}

Once the simulation reaches the final state, the relevant physical observables must be extracted through quantum measurements. While these observables can be computed straightforwardly from the state amplitudes in a classical simulation, as in \cref{eq:expectation_measure}, such amplitudes are not directly accessible on a quantum computer. Instead, expectation values must be reconstructed from measurements, and different classes of observables generally require different measurement strategies.

\textbf{1. Transverse momentum}. Extraction of physical observables such as transverse momentum is straightforward in our setup. As demonstrated in our earlier work~\cite{Qian:2024gph}, it can be computed by direct measurement in the computational momentum basis.

\textbf{2. Cross sections}.
For the convenience of quantum implementation, let us write the cross sections defined in~\cref{eq:crss_defs} in terms of the inner product between the initial and final states.
For the total cross section,
        \begin{align}  \label{eq:cross_tot_inner}
        \begin{split}
        \frac{\diff\sigma_{tot}}{\diff^2 b} 
        =& \langle \sum_{f} |\psi^f_{out}-\psi_{in}^f|^2\rangle_\mathrm{event}\\
    =&\langle \sum_{f} \left(|\psi^f_{out}|^2+|\psi_{in}^f|^2
    -2   \mathrm{Re} (\psi^{f *}_{out}\cdot \psi_{in}^f) \right)
    \rangle_\mathrm{event}\\
       =  &2\left(1 - 
        \langle
        \mathrm{Re}\braket{\psi_{in}|\psi_{out}}
        \rangle_\mathrm{event}
        \right)\;.
        \end{split}
    \end{align}
Here, $\psi^f_{out}= \braket{\phi_f|\psi_{out}}$, $\psi^f_{in}= \braket{\phi_f|\psi_{in}}$, and $\sum_f$ is the summation over the full basis space.  

\begin{figure}
    \centering
    \begin{subfigure}{0.90\textwidth}
        \centering
        \begin{quantikz}
            \lstick{ancilla: $\ket{0}$} & \gate{H} & \qw           & \ctrl{1}        & \gate{H} & \meter{} \\
            \lstick{$\ket{0}$}          & \qwbundle{N_q} & \gate{U_{q\bar{q}}} & \gate{U_{x^+}} & \qw      & \qw
        \end{quantikz}
        \subcaption{Hadamard test circuit for $\mathrm{Re}\braket{\psi_{\mathrm{in}}|\psi_{\mathrm{out}}} = P(0)-P(1)$.}
        \label{fig:qc-xsection-real}
    \end{subfigure}

    \vspace{1em}

    \begin{subfigure}{0.90\textwidth}
        \centering
        \begin{quantikz}
            \lstick{ancilla: $\ket{0}$} & \gate{H}       & \gate{S^\dagger} & \ctrl{1}        & \gate{H} & \meter{} \\
            \lstick{$\ket{0}$}          & \qwbundle{N_q} & \gate{U_{q\bar{q}}}    & \gate{U_{x^+}} & \qw      & \qw
        \end{quantikz}
        \subcaption{Hadamard test circuit for $\mathrm{Im}\braket{\psi_{\mathrm{in}}|\psi_{\mathrm{out}}} = P(0)-P(1)$.}
        \label{fig:qc-xsection-imag}
    \end{subfigure}
    \caption{Quantum circuits for computing the real and imaginary overlaps in evaluating the cross sections $\mathrm{d}\sigma_{\mathrm{total}}/\mathrm{d}^2b$ and $\mathrm{d}\sigma_{\mathrm{el}}/\mathrm{d}^2b$.}
    \label{fig:qc-total-xsection}
\end{figure}
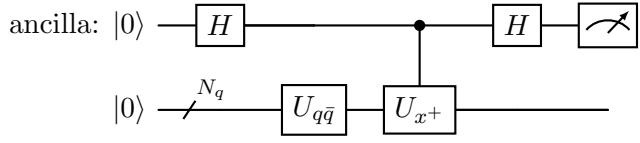
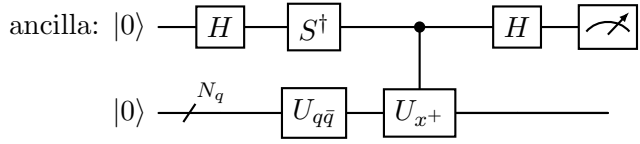

For the elastic cross section, the event average is carried out inside the square,
\begin{align}
    \begin{split}
    \frac{\diff\sigma_{el}}{\diff^2 b} 
    =&\sum_{f} |\langle\psi^f_{out}-\psi_{in}^f\rangle_{\text{event}} |^2\\
    =&\sum_{f} |\langle\psi^f_{out}\rangle_{\text{event}} |^2 + |\psi_{in}^f|^2
    -\langle 2 \mathrm{Re} (\psi^{f*}_{out}\cdot \psi_{in}^f ) \rangle_{\text{event}}\\
      =&\braket{\bar\psi_{out}|\bar\psi_{out} }+1 -2 \mathrm{Re} \braket{\bar\psi_{out}|\psi_{in} }\;,
\end{split}
\end{align}
in which we have defined the event-averaged out state for convenience $\ket{\bar\psi_{out}}\equiv \langle\ket{\psi_{out}}\rangle_{\text{event}}$.
As shown in \cref{eq:crss_eikonal_qqbar_full}, the event-averaged $S$ matrix in the MV model is proportional to the identity in the transverse-position space. It follows that $\langle\bar\psi_\text{out}|\bar\psi_\text{out}\rangle= |\langle\psi_\text{in}|\bar\psi_\text{out}\rangle|^2$, and the elastic cross section simplifies to
\begin{align}\label{eq:cross_el_inner}
\begin{split}
     \frac{\diff\sigma_{el}}{\diff^2 b} 
     &= 
    \lvert
    \braket{ \psi_{in}|\bar \psi_{out}}\rvert^2
    +1 -2 \mathrm{Re} \braket{\bar\psi_{out}|\psi_{in} }\\
     &= 
    \lvert\langle
    \braket{\psi_{in}| \psi_{out}}-1
    \rangle_{\text{event}} \rvert^2\\
    &= 
\lvert\langle
\mathrm{Re} \braket{\psi_{in}|\psi_{out}}-1
\rangle_{\text{event}} \rvert^2
+
 \lvert\langle
\mathrm{Im} \braket{\psi_{in}|\psi_{out}}
\rangle_{\text{event}} \rvert^2
    \;.
\end{split}
\end{align}
With \cref{eq:cross_tot_inner,eq:cross_el_inner}, the summation over the final-state phase space simplifies into a single overlap $\braket{\psi_{in}|\psi_{out}}$ and one no longer needs to enumerate the computational basis states explicitly. One can use the quantum circuits in \cref{fig:qc-total-xsection} for evaluating both the real and imaginary overlaps to extract the total and elastic cross sections, 
requiring only $N_q + 1$ qubits.

\textbf{3. Hadron overlap probability}: To compute $|\braket{\psi_h|\psi(x^+)}|^2$ between the evolved state $\ket{\psi(x^+)}$ and a hadron wave function $\ket{\psi_h}$, in general one can use the standard SWAP test~\cite{Buhrman:2001rma} with two Hadamard gates and one control-SWAP gate; see~\cref{fig:qc-swap}. Measurement on the ancilla directly gives the desired hadron overlap probability: $P(0) = \frac{1}{2}(1 + |\braket{\psi_h|\psi(x^+)}|^2)$. However, this requires $2N_q+1$ for the total number of qubits. A variant that removes the ancilla is the destructive SWAP test~\cite{Garcia-Escartin:2013zvt}; see~\cref{fig:qc-dswap}. A CNOT is applied with register 2 as control and register 1 as target, followed by a Hadamard on register 2, after which both registers are measured. Unlike standard swap test, both registers are consumed upon measurement, but no ancilla or controlled-SWAP gate is needed, reducing the qubit count to $2N_q$.

A better approach would be to utilize variational ansatzes such as the variational quantum eigensolver (VQE)~\cite{Peruzzo:2013bzg} and quantum approximate optimization algorithm~\cite{Farhi:2014ych} to provide an accurate description of the hadron state with a shallow circuit, which is discussed in Ref.~\cite{Kreshchuk:2020aiq,Qian:2021jxp}. Suppose the unitary $U_h$ is available for preparing the hadron, i.e., $\ket{\psi_h}=U_h\ket{0}^{\otimes N_q}$, one can perform an inversion test~\cite{Coles:2018vmk} and measurement on all qubits yield the probability for the overlap; see \cref{fig:qc-inversion}. The advantage is that there will be no extra qubits and is ideal when explicit circuit for both states are available. 

Alternatively, one can always compute the overlap $|\braket{\psi_h|\psi(x^+)}|^2$ using the Hadamard test~\cite{Aharonov:2006dus} to extract the real and imaginary parts in two separate runs using a single ancilla and a controlled-$U_h^\dagger$ operation; see~\cref{fig:qc-hadamard}. Since the overlap is an off-diagonal matrix element, both state preparations must be controlled: the register starts in $\ket{0}^{\otimes N_q}$, and the ancilla controls $U_{q\bar q}$, $U_{x^+}$, and $U_h^\dagger$, where $\ket{\psi(x^+)} = U_{x^+}U_{q\bar q}\ket{0}^{\otimes N_q}$ and $\ket{\psi_h} = U_h\ket{0}^{\otimes N_q}$. The first run gives $P(0) = \frac{1}{2}(1 + \mathrm{Re}\braket{\psi_h|\psi(x^+)})$, and inserting an $S^\dagger$ gate on the ancilla before the final Hadamard in the second run gives $P(0) = \frac{1}{2}(1 + \mathrm{Im}\braket{\psi_h|\psi(x^+)})$, together yielding the full complex overlap at the cost of $N_q + 1$ qubits 
and a deeper circuit due to the controlled evolution. The overlap probability then follows as $|\braket{\psi_h|\psi(x^+)}|^2 = (\mathrm{Re}\braket{\psi_h|\psi(x^+)})^2 + (\mathrm{Im}\braket{\psi_h|\psi(x^+)})^2$  evaluated for each event before averaging.

To model the hadron state, one may consider a simplified ring-state ansatz in which the quark-antiquark pair forms a circular state in relative transverse momentum space, retaining only the momentum component at the half-width
$\kappa$ of the original Gaussian-like wavefunction \cref{eq:phinm}, 
\begin{align}\label{eq:psi_s}
    \ket{\psi_\mathrm{h}} =  \frac{1}{\sqrt{N_c}}\sum_{c=1}^{N_c}
    \mathcal{C}\sum_{k^x,k^y}\delta_{\kappa/d_p,\sqrt{(k^x)^2+(k^y)^2}}
    \ket{q(\vec k_\perp, c) \bar q(-\vec k_\perp, c)}\;,
\end{align}
where $\mathcal{C}$ is the normalization factor.
To implement the Kronecker delta $\delta_{\kappa/d_p,\sqrt{(k^x)^2+(k^y)^2}}$ on the lattice, both $\kappa/d_p$ and $\sqrt{(k^x)^2+(k^y)^2}$ are rounded to the nearest integer before enforcing the equality condition. On the circuit, this is a generalized W state, which can be constructed analagously as in Eq.~\eqref{eq:wstate_angles}.

Finally, to obtain the physical results to account for the fluctuation in the background medium, we repeat the extraction of physical observables over $N_\mathrm{event}$ events for multiple background configurations. This process closely resembles the physical scenario in the high-energy physics experiments where multiple events happen.

\newlength{\circuitheight}
\setlength{\circuitheight}{3cm}

\begin{figure}[t]
\centering
\begin{subfigure}{0.45\textwidth}
    \centering
    \begin{minipage}[c][\circuitheight][c]{\textwidth}
        \centering
        \begin{adjustbox}{max width=\linewidth}
        \begin{quantikz}
            \lstick{ancilla $\colon \ket{0}$} & \gate{H} & \ctrl{1} & \gate{H} & \meter{} & \qw \\
            \lstick{reg.\,2 $\colon \ket{\psi(x^+)}$} & \qwbundle{N_q} & \swap{1} & \qw & \qw & \qw \\
            \lstick{reg.\,1 $\colon \ket{\psi_h}$} & \qwbundle{N_q} & \targX{} & \qw & \qw & \qw
        \end{quantikz}
         \end{adjustbox}
    \end{minipage}
    \subcaption{SWAP test circuit.}
    \label{fig:qc-swap}
\end{subfigure}
\hfill
\begin{subfigure}{0.45\textwidth}
    \centering
    \begin{minipage}[c][\circuitheight][c]{\textwidth}
        \centering
        \begin{adjustbox}{max width=\linewidth}
        \begin{quantikz}
            \lstick{reg.\,2 $\colon \ket{\psi(x^+)}$} & \qwbundle{N_q} & \ctrl{1} & \gate{H} & \meter{} & \qw \\
            \lstick{reg.\,1 $\colon \ket{\psi_h}$} & \qwbundle{N_q} & \targ{} & \qw & \meter{} & \qw
        \end{quantikz}
        \end{adjustbox}
    \end{minipage}
    \subcaption{Destructive SWAP test circuit.}
    \label{fig:qc-dswap}
\end{subfigure}

\vspace{1.5em}

\begin{subfigure}{0.38\textwidth}
    \centering
    \begin{minipage}[c][\circuitheight][c]{\textwidth}
        \centering
         \begin{adjustbox}{max width=\linewidth}
        \begin{quantikz}
            \lstick{reg.\,1 $\colon \ket{\psi(x^+)}$} & \qwbundle{N_q} & \gate{U_h^\dagger} & \qw & \meter{} & \qw
        \end{quantikz}
        \end{adjustbox}
    \end{minipage}
    \subcaption{Inversion test circuit.}
    \label{fig:qc-inversion}
\end{subfigure}
\hfill
\begin{subfigure}{0.58\textwidth}
    \centering
    \begin{minipage}[c][\circuitheight][c]{\textwidth}
        \centering
        \begin{adjustbox}{max width=\linewidth}
        \begin{quantikz}
            \lstick{ancilla $\colon \ket{0}$} & \gate{H} & \ctrl{1} & \ctrl{1} & \ctrl{1} & \gate{S^\dagger}\gategroup[1,steps=1,style={dashed,rounded corners,inner xsep=2pt,inner ysep=2pt},
                    label style={label position=above,anchor=south,yshift=-0.1cm}]{{\scriptsize 2nd run}}
            & \gate{H} & \meter{} \\
            \lstick{reg.\,1  $\colon \ket{0}$} & \qwbundle{N_q} & \gate{U_{q\bar q}} & \gate{U_{x^+}} & \gate{U_h^\dagger} & \qw & \qw & \qw
        \end{quantikz}
        \end{adjustbox}
    \end{minipage}
    \subcaption{Hadamard test circuit.}
    \label{fig:qc-hadamard}
\end{subfigure}

\caption{Quantum circuits for computing the hadron overlap probability 
$|\braket{\psi_h|\psi(x^+)}|^2$.}
\label{fig:qc-hadron-prob}
\end{figure}
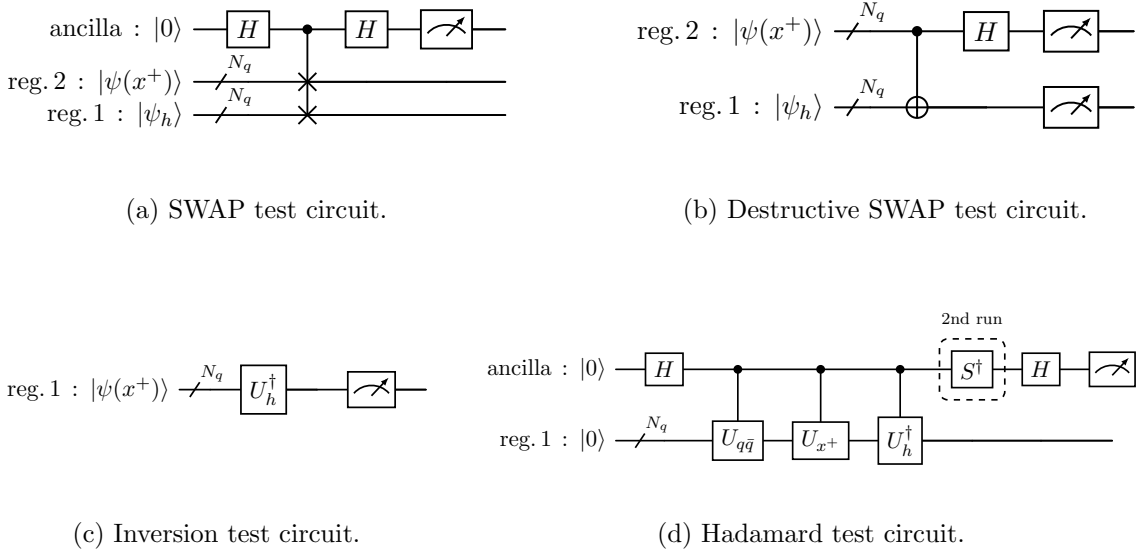

\section{Results\label{sec:results}}

Using the method formulated above, we perform numerical calculations using both classical and quantum circuit simulation approaches. We consider two configurations of the initial dipole state: the single-momentum state, and the photon state. We focus on the evolution of the momentum distribution of the dipole, the cross section, the hadron production rate, and its quantum entanglement.

For the main results of the paper, we employ classical simulation using direct matrix exponentiation. 
We use the following parameters throughout this section for the classical simulations.
The transverse lattice size is $L_\perp = 12~\text{GeV}^{-1}$ with $N_\perp = 4$ sites and $N_c=3$.
For the photon initial state, we take the longitudinal modes $K=8$. For the background field describing the nuclear target, the charge density is $g^2\tilde\mu=0.8~\text{GeV}^{3/2}$ with IR regulator $m_g$=0.3~\text{GeV}.
The extension of the field is $L_{\eta,p}=6~\GeV^{-1}$ for proton and 
$L_{\eta,Au}=35~\GeV^{-1}$ for Au nuclei, each partitioned into $N_\eta=16$ layers along the longitudinal direction. Since we project the final states to charmonium states, we choose charm quark mass $m_f = 1.27~\text{GeV}$ and $Q_f=2/3$. 
We work in the eikonal limit, $P^+ = \infty$.
Results are averaged over $N_\mathrm{event}=128$ configurations, with the standard error shown as the uncertainty bar.

To demonstrate the validity of the quantum circuit, we also provide the quantum circuit results on a smaller transverse lattice of $N_\perp =2$ and $N_c=2$. This requires a total of 65 qubits (with one ancilla qubit) with the direct encoding scheme used in our paper. For the parameters, we used $L_\perp=32 \GeV^{-1}, g^2\tilde\mu=0.4 \GeV^{3/2}, m_g=0.1 \GeV, N_\eta = 16$. The quantum simulation are performed using \texttt{Qiskit}~\cite{QiskitCommunity2017} and the results are averaged over $N_\mathrm{event}=5$ configurations.

\subsection{Momentum distribution and broadening}
\begin{figure}
    \centering
    \includegraphics[width=0.45\linewidth]{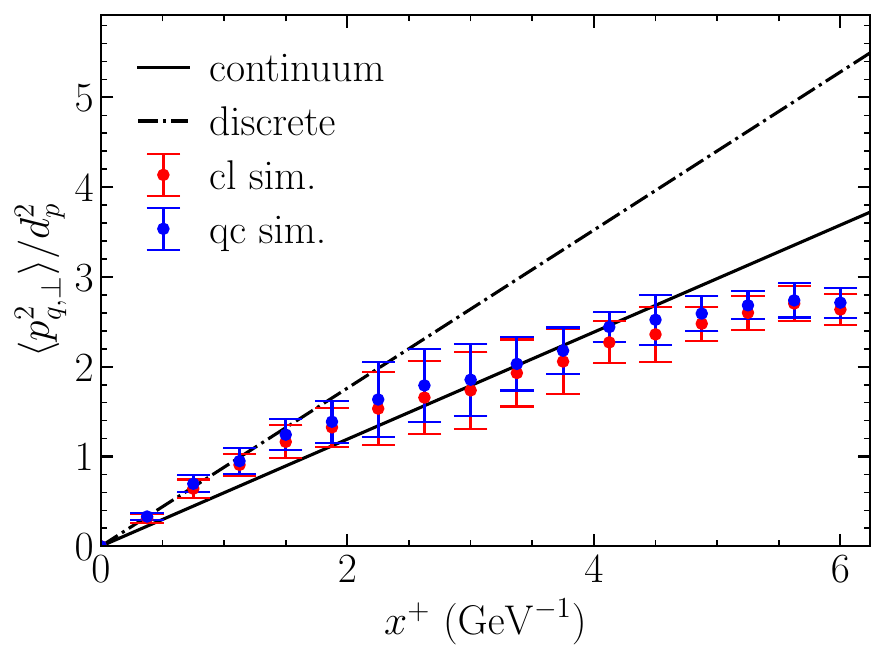}\quad\quad
    \includegraphics[width=0.45\linewidth]{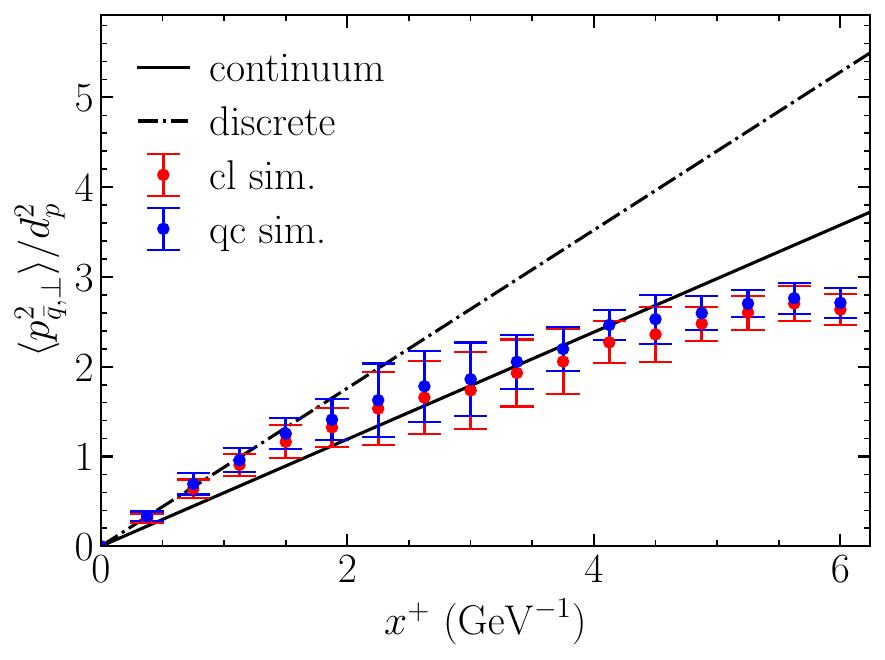}
    \caption{
    Momentum broadening of quark and anti-quark of the initial single-momentum dipole for the proton using classical and quantum circuit simulations.
    }
    \label{fig:qc-res-psquare}
\end{figure}

We begin by validating the quantum-circuit implementation against the
classical benchmark. 
\Cref{fig:qc-res-psquare} shows the transverse momentum
square $\braket{p^2_\perp}/d_p^2$ of the quark (left) and the antiquark
(right) as a function of the light-front time $x^+$, for the initial
single-momentum dipole in the nuclear field. 
The classical (cl sim.) and quantum-circuit (qc sim.) results agree within event uncertainties over the entire evolution. The quark and antiquark results are identical, as expected from the symmetry of the color-singlet dipole under $q\leftrightarrow\bar q$.

To interpret the time dependence, we compare the simulations with the
analytical results derived in \cite{Barata:2022wim,Barata:2023clv}. 
In the continuum, the broadening reads
\begin{align}
\langle p_\perp^2 (x^+) \rangle_{\rm continuum} &= \frac{C_F \, (g^2\tilde \mu)^2 x^+}{4\pi} \left[\ln\!\left(1 + \frac{\pi^2}{a_\perp^2 m_g^2}\right) - \frac{1}{1 + a_\perp^2 m_g^2/\pi^2}\right]  \,, \label{eq:p2_ana}
\end{align}
while on the discrete transverse lattice with $2N_\perp$ sites in each
direction it becomes
\begin{align}
\langle p_\perp^2(x^+)\rangle_{\rm discrete} &= \frac{C_F \, (g^2\tilde \mu)^2 x^+}{4\pi^2}
\sum_{k_x=-N_\perp}^{N_\perp-1}
\sum_{k_y=-N_\perp}^{N_\perp-1}
\frac{k_x^2 + k_y^2}{\left(\dfrac{N_\perp^2 a_\perp^2 m_g^2}{\pi^2} + k_x^2 + k_y^2\right)^{\!2}} \,, \label{eq:p2_lat}
\end{align}
which reduces to \cref{eq:p2_ana} for $N_\perp\to\infty$. Both analytical
results grow linearly in $x^+$, reflecting the diffusive transfer of
transverse momentum from the stochastic color field.
At later times, the broadening in both simulations deviates from the linear growth of \cref{eq:p2_lat} and saturates. 
This is a finite-basis effect. The linear growth of the analytical results assumes that the momentum distribution can spread without bound, whereas in the simulations it is confined to the $(2N_\perp)^2$ transverse modes with periodic boundary conditions. 
Once the distribution reaches the edge of the momentum box, kicks across the boundary wrap around to the opposite side and no longer increase $\braket{p_\perp^2}$. 
At asymptotically late times, the distribution becomes uniform over all modes, for which $\braket{p_\perp^2}_{\rm asy}/d_p^2=2N_\perp^2/3\approx 2.67$ for $N_\perp=2$ \cite{Li:2023jeh}. 
Neither analytical result accounts for this bound, so both are valid only at early times. 
In particular, the crossing of the simulations with the continuum result near $x^+\approx 4~{\rm GeV}^{-1}$ is accidental and should not be read as
agreement.

\begin{figure}
    \centering
    \begin{subfigure}{\linewidth}
    \includegraphics[width=\linewidth]{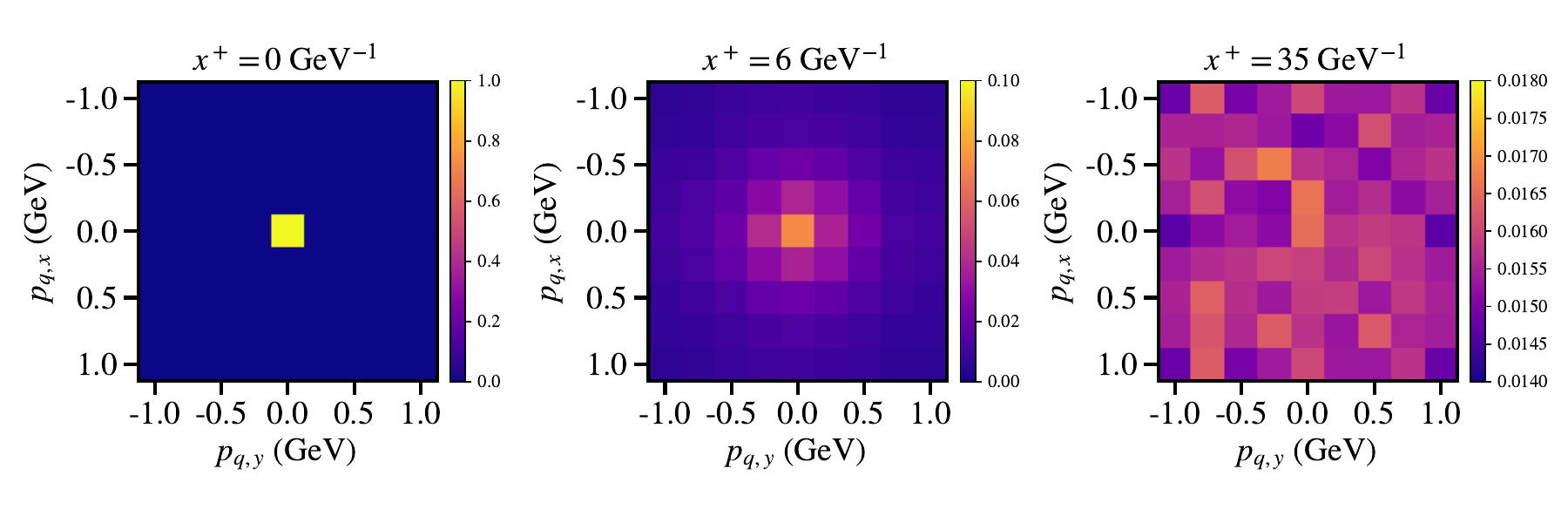}
    \caption{quark momentum distribution\label{fig:qqbar_pq}} 
    \end{subfigure}
    \begin{subfigure}{\linewidth}
    \includegraphics[width=\linewidth]{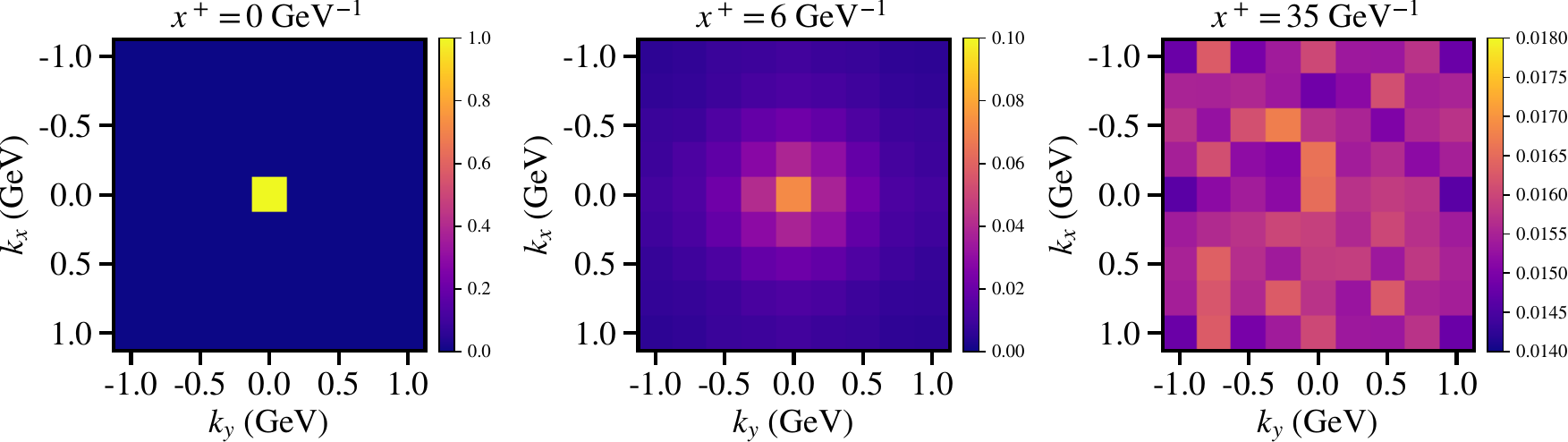}
    \caption{relative momentum distribution \label{fig:qqbar_kt}}
    \end{subfigure}
    \begin{subfigure}{\linewidth}
    \includegraphics[width=\linewidth]{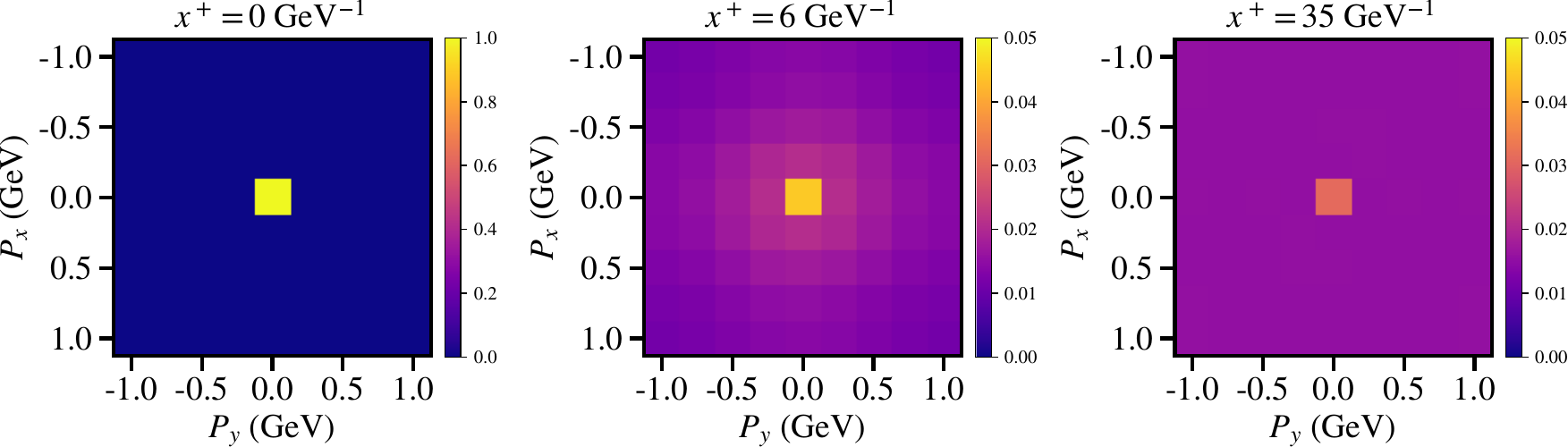}
    \caption{center-of-mass momentum distribution \label{fig:qqbar_PCM}}
    \end{subfigure}
    \caption{Momentum distribution of the initially single-momentum dipole at selected light-front time. 
    }
    \label{fig:qqbar-rel-t}
\end{figure}

Next, \cref{fig:qqbar-rel-t} shows the evolution of the momentum distributions of the $q\bar{q}$, starting from the single-momentum dipole in \cref{eq:dipole_LFWF}.
The top row \cref{fig:qqbar_pq} shows the quark momentum $\vec p_{q,\perp}$. 
As the dipole propagates through the medium, it exchanges transverse momentum with the background field, and the probability spreads from the initial mode to its neighbors.
At $x^+=6~\GeV^{-1}$ the distribution is broadened but still peaked at the origin. At $x^+=35~\GeV^{-1}$ it has become nearly uniform, an asymptotic state on the finite basis. 
The middle row \cref{fig:qqbar_kt} shows the boost-invariant relative momentum $\vec k_\perp$ defined in \cref{eq:CMP_k}, whose distribution closely follows that of the quark at all times.

The bottom row \cref{fig:qqbar_PCM} shows the C.M. transverse momentum $\vec P_\perp$.
At $x^+=6~{\rm GeV}^{-1}$ its distribution extends further from the origin than that of the quark, i.e., the C.M. momentum broadens faster. This is expected because the single-momentum dipole is fully delocalized in relative position. The quark and the antiquark are therefore typically at different positions and receive independent kicks from the color field, so that their contributions add, $\braket{P_\perp^2}\approx\braket{p_{q,\perp}^2}
+\braket{p_{\bar q,\perp}^2}=2\braket{p_{q,\perp}^2}$. 
Unlike the quark distribution, the C.M. distribution retains a peak at $\vec P_\perp=0$ even at $x^+=35~{\rm GeV}^{-1}$. This peak comes from the component of the state in which the quark and the antiquark occupy the same transverse site. 
There the color-singlet pair forms a dipole of zero size, which does not interact with the background field (color transparency), so its C.M. momentum is unchanged. 
This component carries a probability of $1/(2N_\perp)^2\approx 0.016$ for $N_\perp=4$, which adds to the roughly uniform contribution of $\approx 1/(2N_\perp)^2$ from the rest of the state.
The $\vec P_\perp=0$ mode is therefore populated at about twice the uniform
value, consistent with the figure.

\begin{figure}
    \centering
    \begin{subfigure}{\linewidth}
    \includegraphics[width=\linewidth]{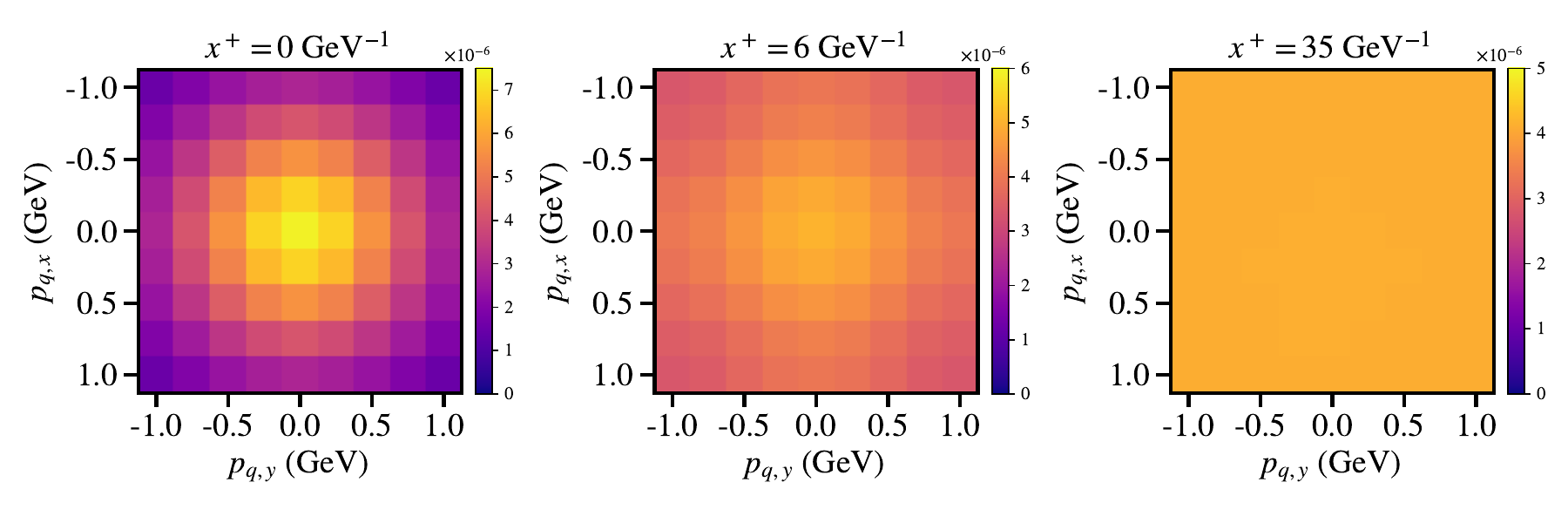}
    \caption{quark momentum distribution \label{fig:photon_pq}}
    \end{subfigure}
    \begin{subfigure}{\linewidth}
    \includegraphics[width=\linewidth]{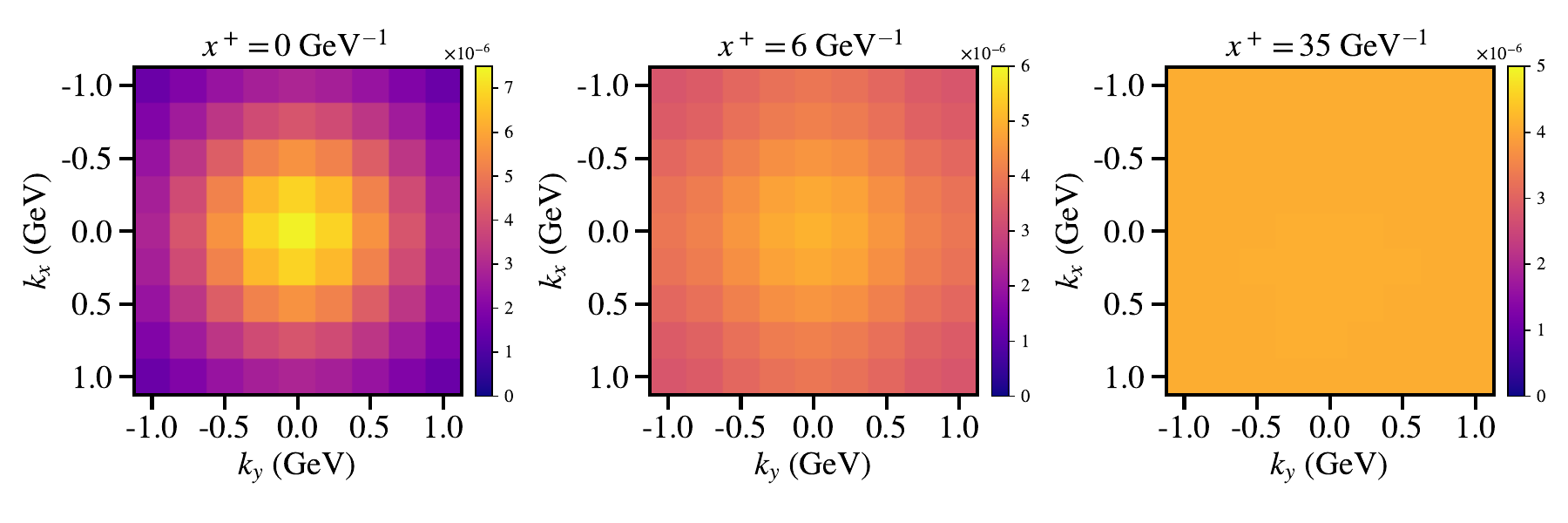}
    \caption{relative momentum distribution \label{fig:photon_kt}}
    \end{subfigure}
    \begin{subfigure}{\linewidth}
    \includegraphics[width=\linewidth]{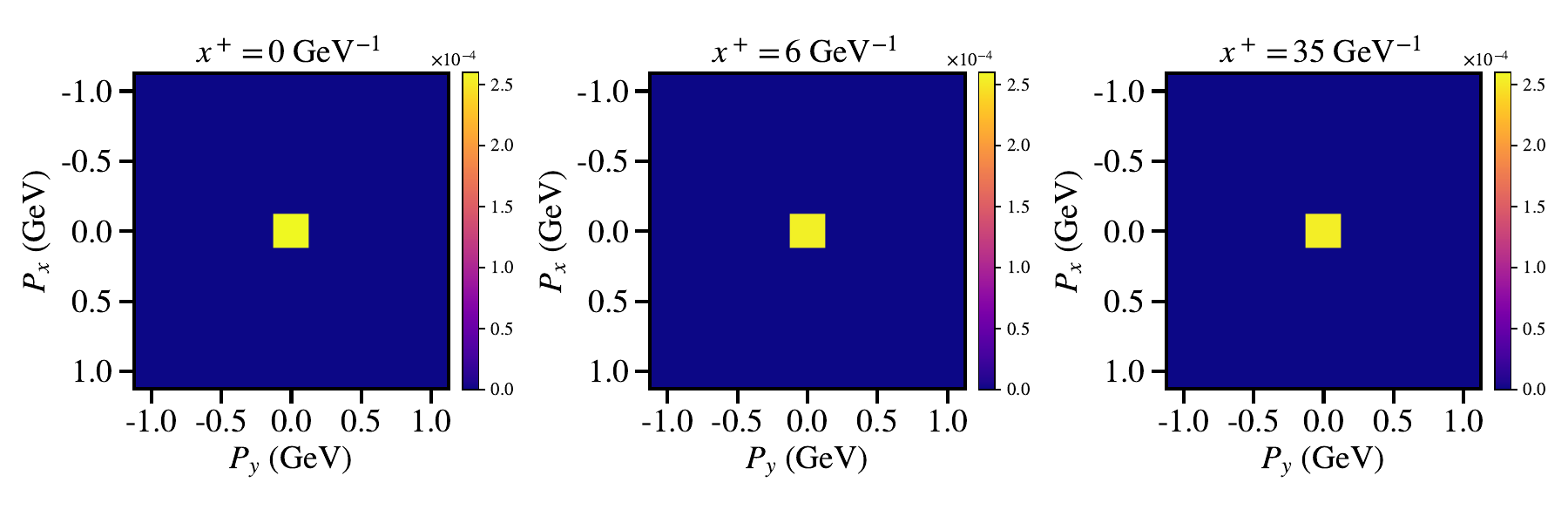}
    \caption{center-of-mass momentum distribution \label{fig:photon_PCM}}
    \end{subfigure}
    \caption{Momentum distribution of the initially virtual photon state at selected light-front time. The initial photon LFWF $\psi_{\uparrow\uparrow/\gamma}^{(m_j=0)}(x, \vec{k}_\perp)$ is with virtuality $Q=1~\text{GeV}$. 
    }
    \label{fig:photon-rel-t}
\end{figure}

We next examine the evolution of the dipole's transverse-momentum distribution when the initial state is a photon. The corresponding light-front wavefunction of the $q\bar{q}$ pair is given in \cref{eq:photon_LFWF}, and for simplicity, we consider the $m_j=0$ component here. 
The results of this evolution are shown in \cref{fig:photon-rel-t}. 
In contrast to the single-momentum case in \cref{fig:qqbar-rel-t}, the photon-initiated distribution initially exhibits a wave-packet-like profile, peaked at $\vec 0_\perp$. 
This profile broadens over time and, and by $x^+=35~\GeV^{-1}$, becomes uniform over all transverse modes, as for the single-momentum dipole. The two uniform distributions differ only in their overall magnitude. 
The shape of the asymptotic distribution is independent of the initial state, as expected for diffusion on a finite basis.

By comparing \cref{fig:photon_pq} and \cref{fig:photon_kt}, we can see that the broadening of the quark's momentum $\vec p_{q,\perp}$ and that of the relative momentum $\vec k_\perp$ are nearly identical at all times.
This follows from \cref{eq:CMP_k}, which gives $\vec k_\perp=\vec p_{q,\perp}-x\vec P_\perp$. As shown in \cref{fig:photon_PCM}, the C.M. momentum of the photon state remains concentrated at $\vec P_\perp\approx0$ throughout the evolution, with no visible broadening, so that $\vec k_\perp\approx\vec p_{q,\perp}$.
The broadening of the CM momentum, as shown in \cref{fig:photon_PCM}, is not noticeable.
This is because the quark and antiquark in the photon state is highly correlated. As can be estimated from the photon's LFWF in \cref{eq:photon_LFWF_mj0}, the typical separation between the quark and the antiquark is $r_\perp\sim 1/\epsilon=1/\sqrt{x(1-x)Q^2 + m_f^2}\sim 0.73 \GeV^{-1}$, much smaller than transverse lattice spacing $a_\perp\approx3.0~\GeV^{-1}$, the smallest separation resolved by the basis.
At this resolution the quark and the antiquark essentially occupy the same transverse site, so the color-singlet pair is transparent to the background field and its total transverse momentum is unchanged. The single-momentum dipole, as shown in \cref{fig:qqbar_PCM}, by contrast, is fully delocalized in relative position, and its constituents receive independent kicks.

\subsection{Dipole cross sections}
\begin{figure}[htbp]
    \centering
    \begin{subfigure}[b]{\textwidth}
        \centering
        \includegraphics[width=0.45\textwidth]{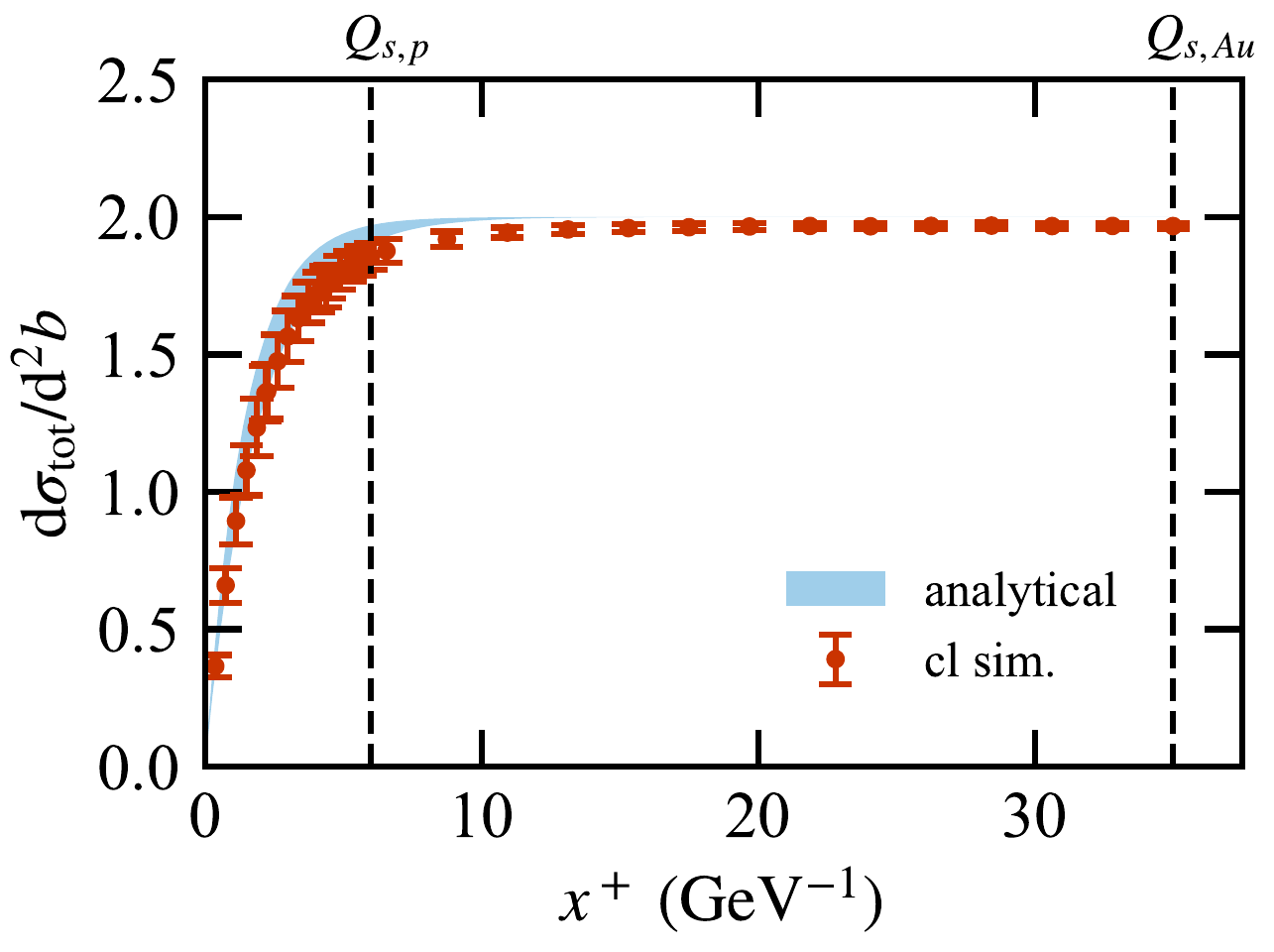}
        \includegraphics[width=0.45\textwidth]{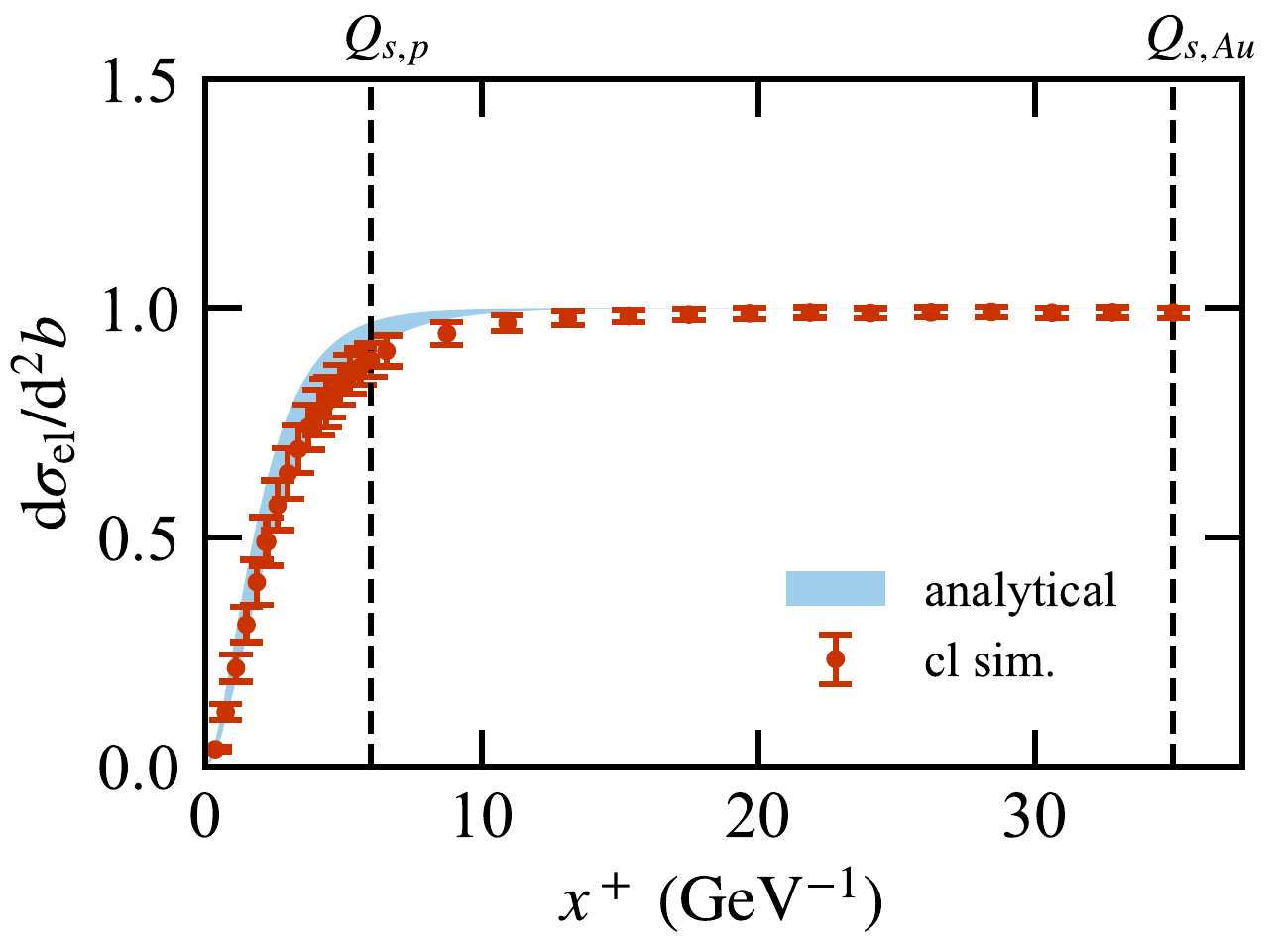}
        \caption{Single momentum dipole \label{fig:crss_dipole}}
    \end{subfigure}
    \hfill
    \begin{subfigure}[b]{\textwidth}
        \centering
        \includegraphics[width=0.45\textwidth]{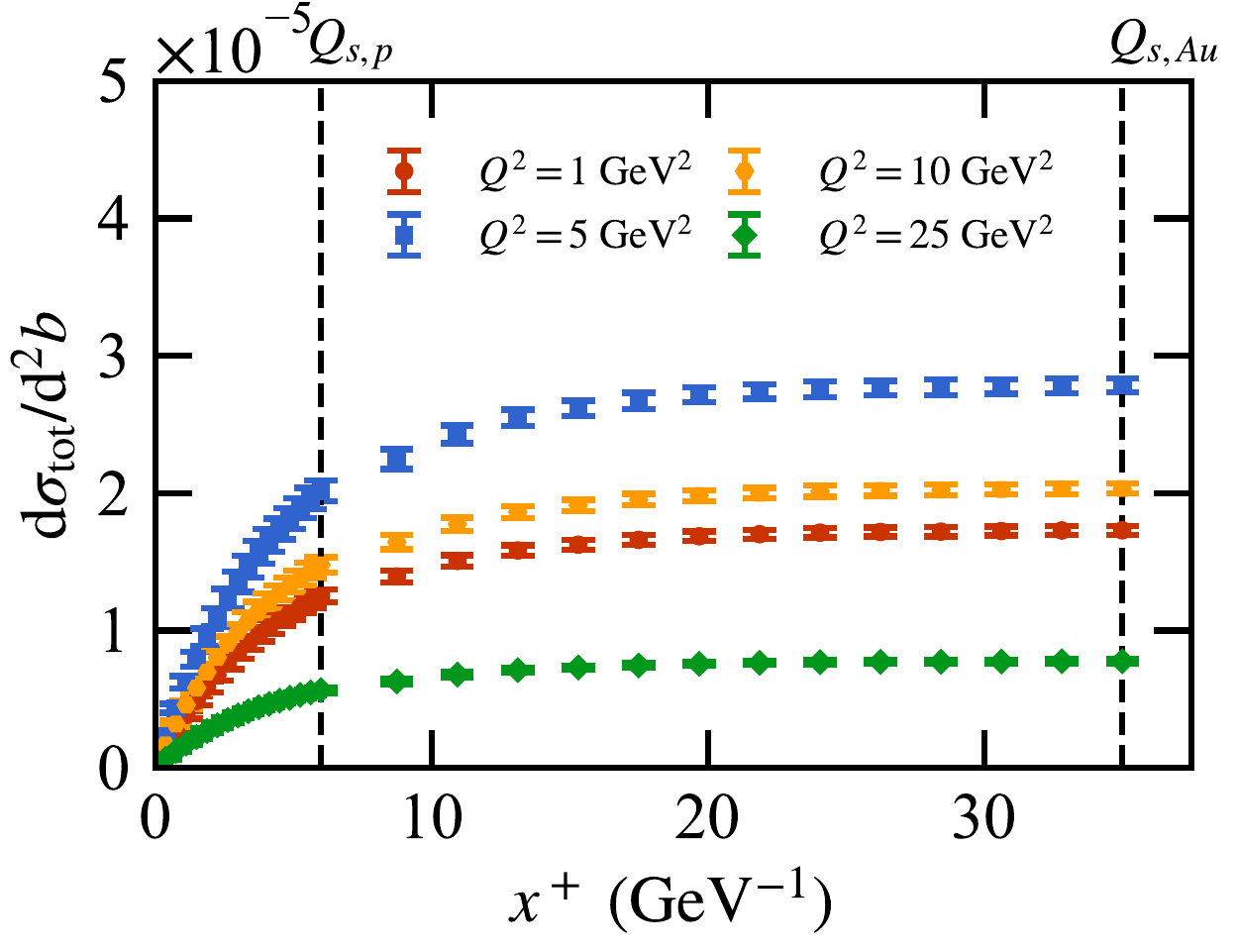}
        \includegraphics[width=0.45\textwidth]{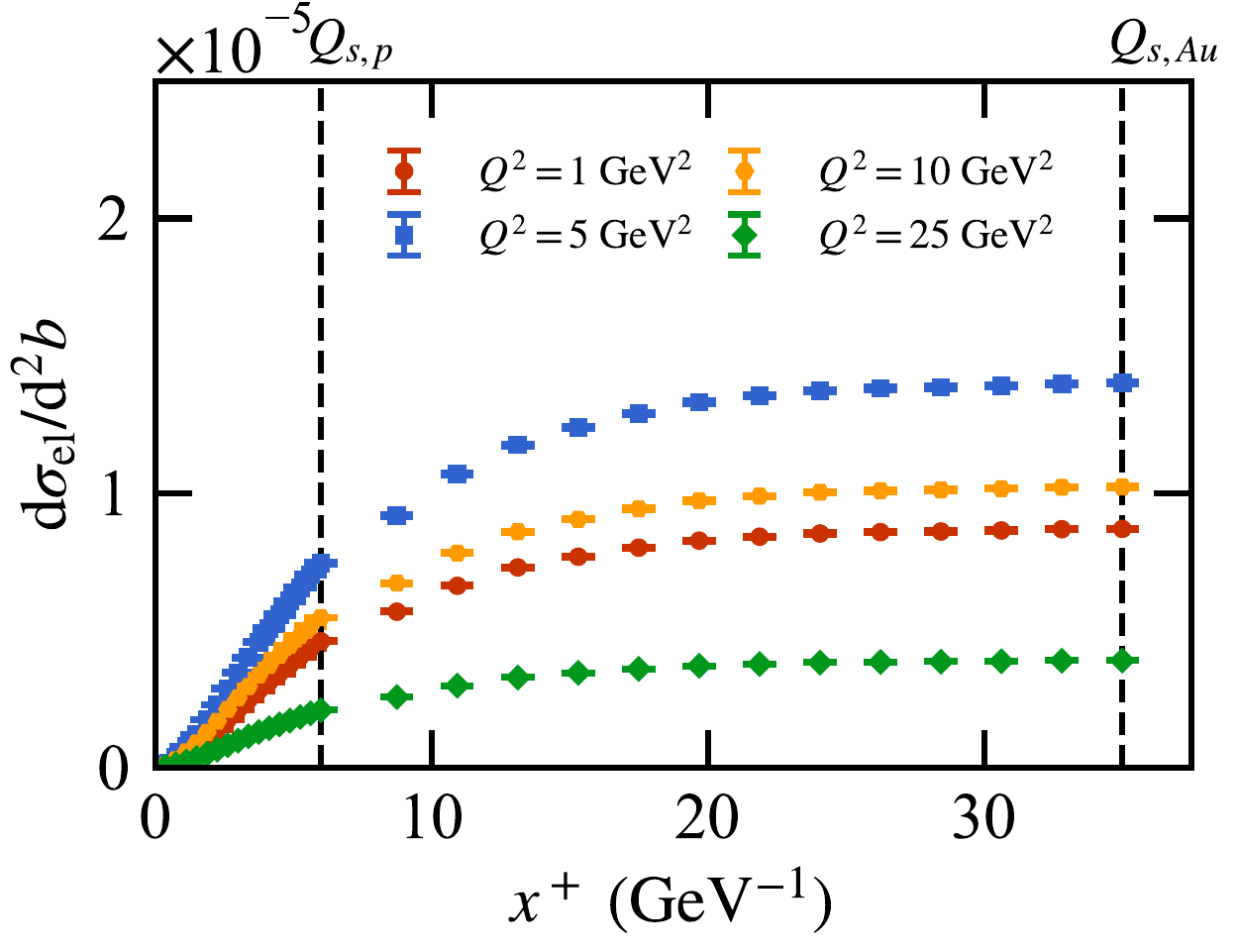}
        \caption{Virtual photon state \label{fig:crss_photon}}
    \end{subfigure}
    \caption{Cross sections of (a) the initial single-momentum dipole and (b) the virtual photon state as a function of light-front time $x^+$. 
    The blue bands show the analytic eikonal predictions given by \cref{eq:crss_eikonal_qqbar_total} and \cref{eq:crss_eikonal_qqbar_elastic}, where the upper and lower bounds are obtained by taking $|\vec{x} - \vec{y}| = L_\perp$ and $L_\perp/2$, respectively.
    }
    \label{fig:qqbar-SU3-total-cross-section}
\end{figure}

We first examine the cross sections of the a single-momentum dipole. 
\Cref{fig:crss_dipole} shows the total and elastic cross sections as functions of the evolution time. 
The color charge density is fixed at $g^2\tilde\mu = 0.8~\text{GeV}^{3/2}$, so that increasing $x^+$ corresponds to a thicker target with a larger saturation scale.
The two vertical dashed lines mark the values of $x^+$ at which the saturation scale matches that of the proton and of the Au nucleus, respectively, and the cross sections at these times give the results for the two targets. 
Both cross sections rise rapidly and approach the black-disk limits at ${\rm d}\sigma_{\rm tot}/{\rm d}^2b=2$ and ${\rm d}\sigma_{\rm el}/{\rm d}^2b=1$. 
The dipole is already close to this limit for the proton and essentially reaches it for the Au nucleus.

The blue bands show the analytical eikonal results from \cref{eq:crss_eikonal_qqbar_full}. 
These are derived for a dipole of fixed transverse size, whereas the quark and the antiquark in the single-momentum dipole are delocalized over the whole transverse plane.
We therefore bracket the analytical result by evaluating it at the largest separation on the periodic lattice, $\lvert\vec r_{\perp,q}-\vec r_{\perp,\bar q}\rvert=L_\perp$, for the upper boundary, and at $L_\perp/2$ for the lower boundary. 
The narrow width of the band indicates that the result is insensitive to the precise separation in this range. 
The classical simulation agrees with the analytical band, with only small deviations.

\Cref{fig:crss_photon} shows the total and elastic cross sections of the photon state for virtualities $Q^2=1$, $5$, $10$, and $25~\GeV^2$, obtained from the same evolution at fixed charge density as for the single-momentum dipole.
The overall magnitude, of order $10^{-5}$, reflects the small norm of the $q\bar q$ component in the photon state in the basis space.
As for the single-momentum dipole, the cross sections rise and saturate, and the results for the Au nucleus are larger than those for the proton. 
At late times, the elastic cross section is half the total one for all $Q^2$, as in the black-disk limit.
The approach to saturation is slower than for the single-momentum dipole.
Since the quark and the antiquark in the photon are closer together in transverse space, a larger part of the state is color transparent, i.e., the effects of the background field on the quark and the antiquark largely cancel for the color-singlet pair.
A thicker target is therefore needed to drive the photon state away from its initial configuration. In the single-momentum dipole, by contrast, the quark and the antiquark are delocalized and interact with the field mostly independently.
The cross sections depend non-monotonically on $Q^2$: they increase from $Q^2=1$ to $5~\GeV^2$ and decrease at larger $Q^2$.

\begin{figure}[htbp]
    \centering
    \begin{subfigure}[b]{0.45\textwidth}
        \centering
        \includegraphics[width=\textwidth]{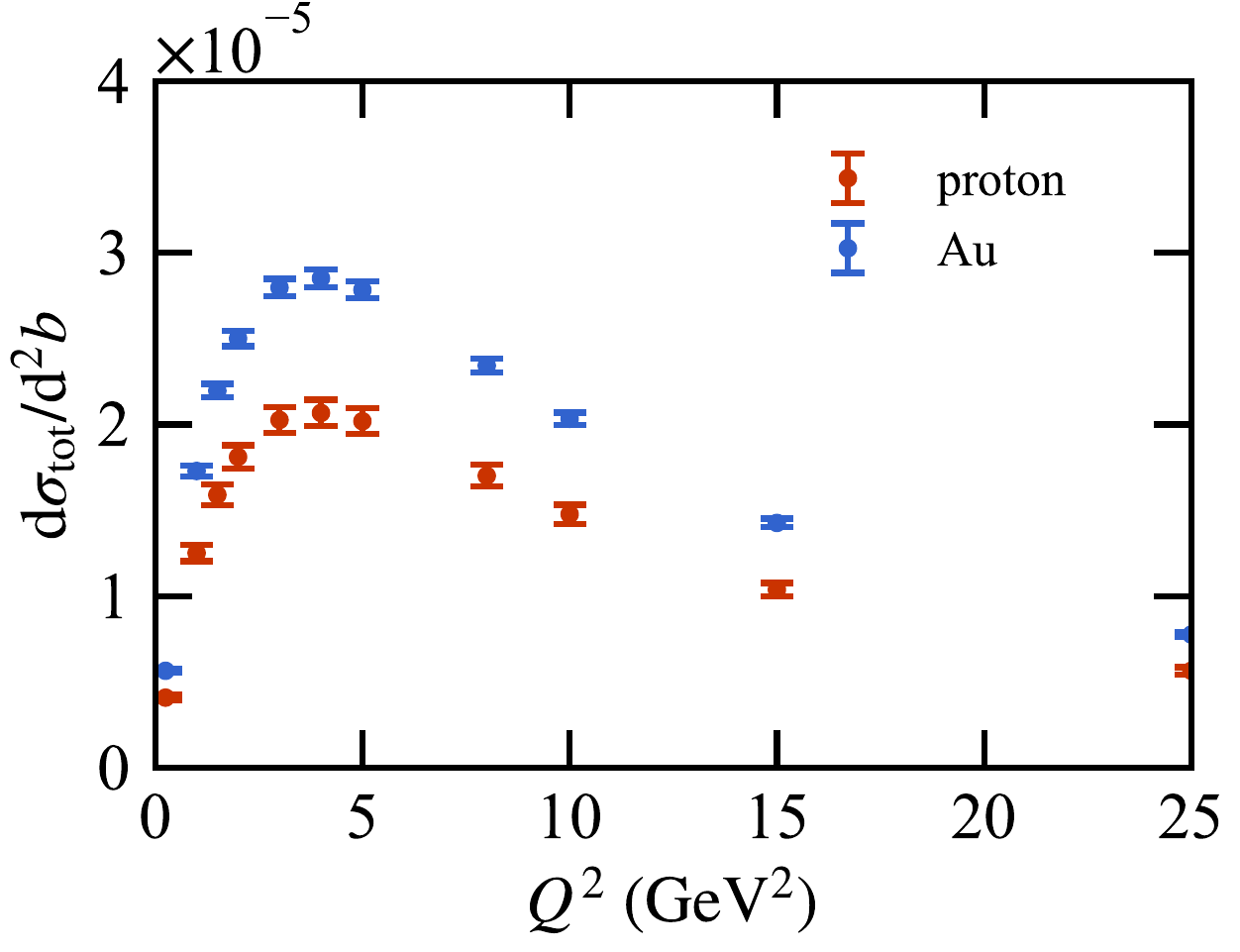}
        \caption{The total cross section\label{fig:photon-cross-section_tot}} 
    \end{subfigure}
    \hfill
    \begin{subfigure}[b]{0.45\textwidth}
        \centering
        \includegraphics[width=\textwidth]{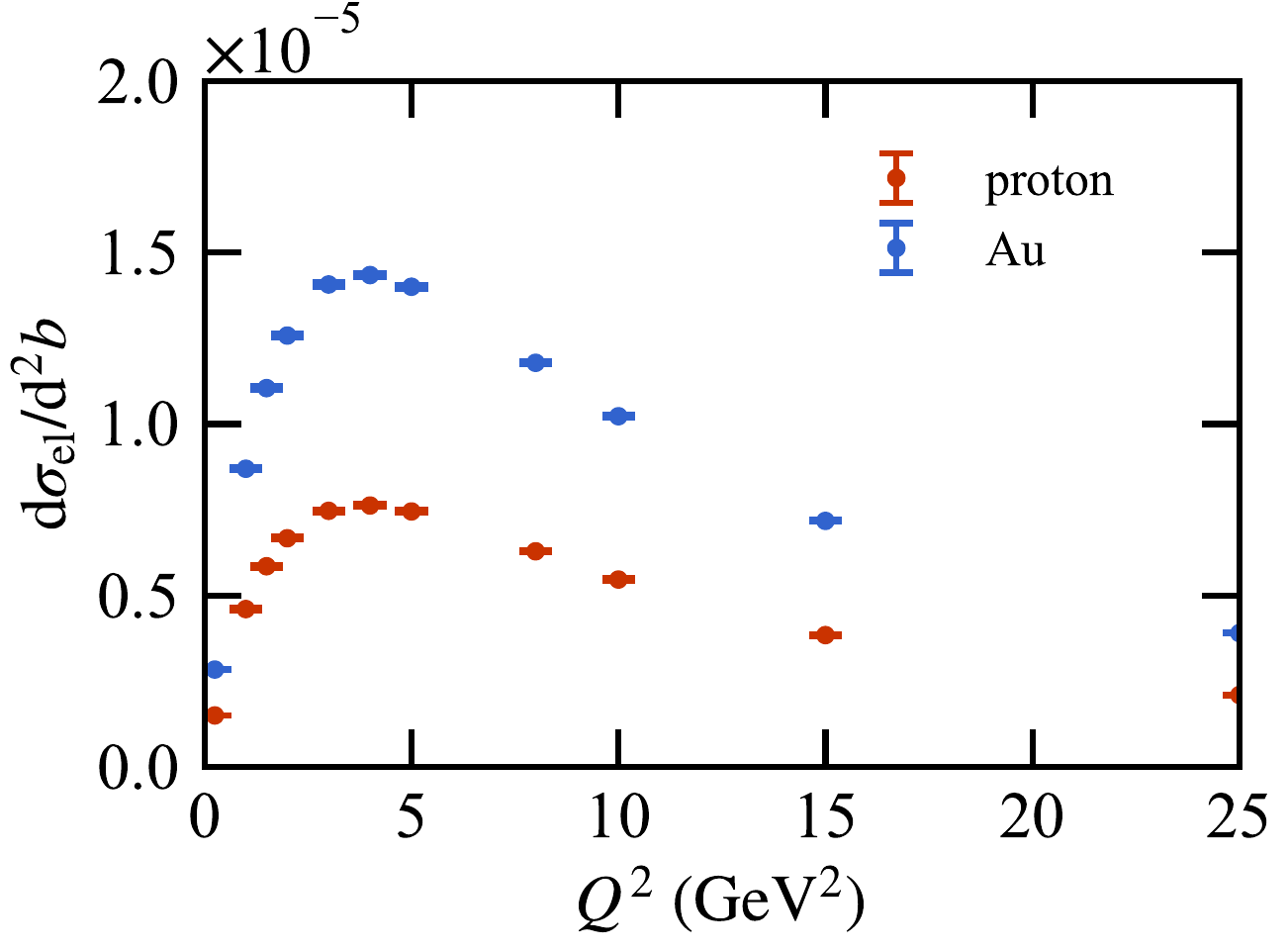}
        \caption{The elastic cross section\label{fig:photon-cross-section_el}}
    \end{subfigure}
    \caption{Cross sections of the photon state as a function of virtuality $Q^2$ for both the proton and Au nucleus.}
    \label{fig:photon-cross-section}
\end{figure}

To further examine the dependence of the photon's cross section on $Q^2$, we present the total and elastic cross sections as functions of $Q^2$ in \cref{fig:photon-cross-section}.
For both targets, the cross section first rises with $Q^2$, reach a maximum around $Q^2\approx 4\GeV^2$, and then decrease, as seen in \cref{fig:photon-cross-section_tot} for the total cross section; the elastic cross section in \cref{fig:photon-cross-section_el} follows the same trend. The Au cross sections are systematically larger than those of the proton, as expected from its larger saturation scale.

\begin{figure}
    \centering
    \includegraphics[width=0.5\textwidth]{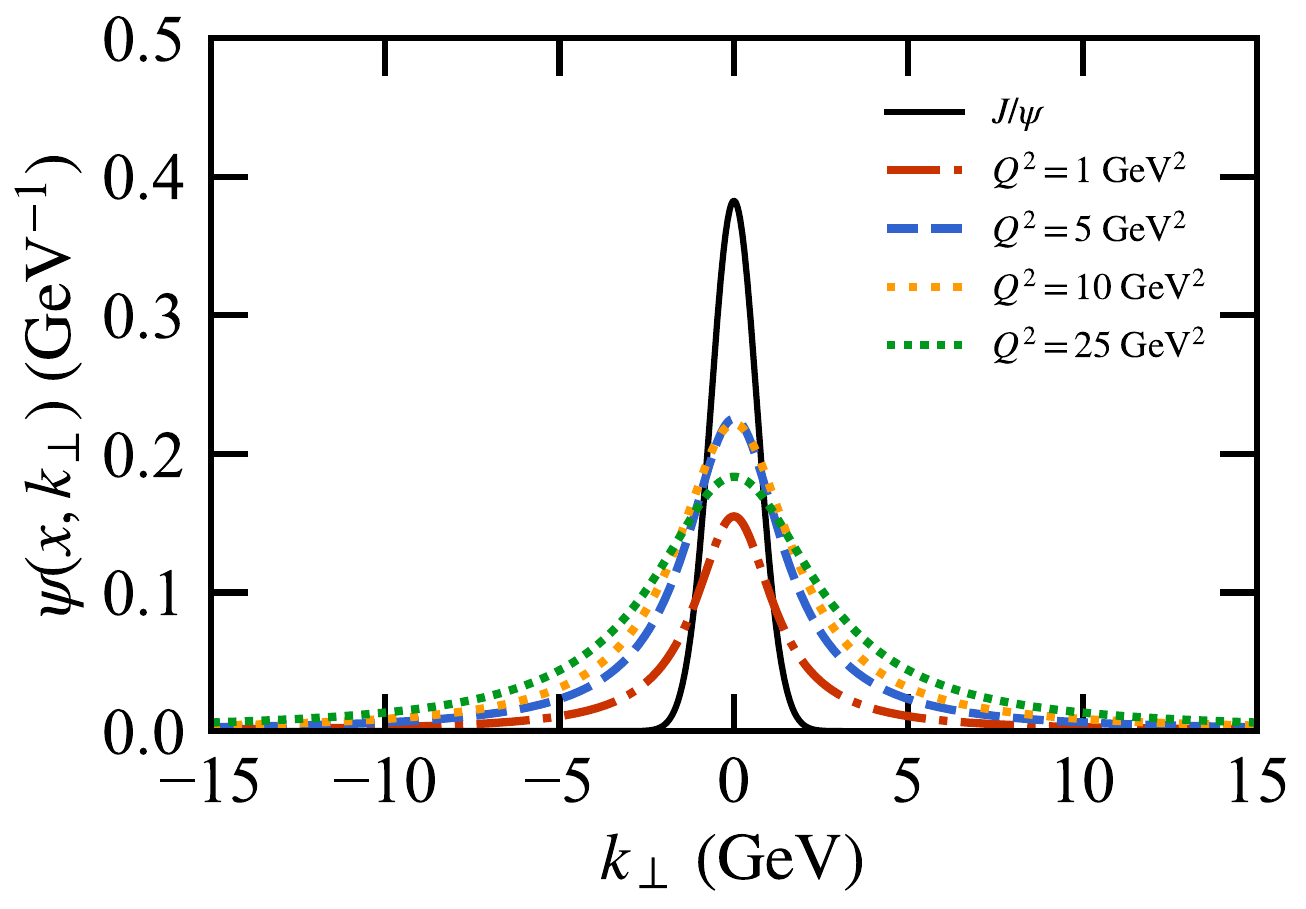}
    \caption{
    The photon LFWF $\psi_{\uparrow\downarrow+\downarrow\uparrow/\gamma}^{m_j=0}$ according to \cref{eq:photon_LFWF_mj0} at $x=0.5$ for several photon virtualities, compared with the $J/\psi$ wavefunction according to \cref{eq:WF_base1_mj0} (scaled down by a factor of 40). The quark mass is $m_q=1.27~\text{GeV}$.
}
    \label{fig:WF_compare}
\end{figure}

The non-monotonic behavior can be understood from the $Q^2$ dependence of the photon wave function, since for a given target the cross section is determined by how strongly the interaction drives the initial $q \bar{q}$ state away from itself, i.e., by its overlap with the evolved state, which approaches a color-momentum-randomized (uniformly distributed) state at late times.
Two competing effects are at play. At smaller $Q^2$, the amplitude of the $q \bar{q}$ component of the photon grows with $Q^2$, as illustrated in \cref{fig:WF_compare}, which enhances the cross section. At larger $Q^2$, the wave function broadens in $k_\perp$ as $Q^2$ increases, corresponding to dipoles of smaller transverse size, $r_\perp\sim 1/\sqrt{x(1-x)Q^2 + m_q^2}$. 
Smaller dipoles are expected to interact more weakly with the target, and the cross section decreases accordingly.

\begin{figure}
    \centering
    \includegraphics[width=0.45\linewidth]{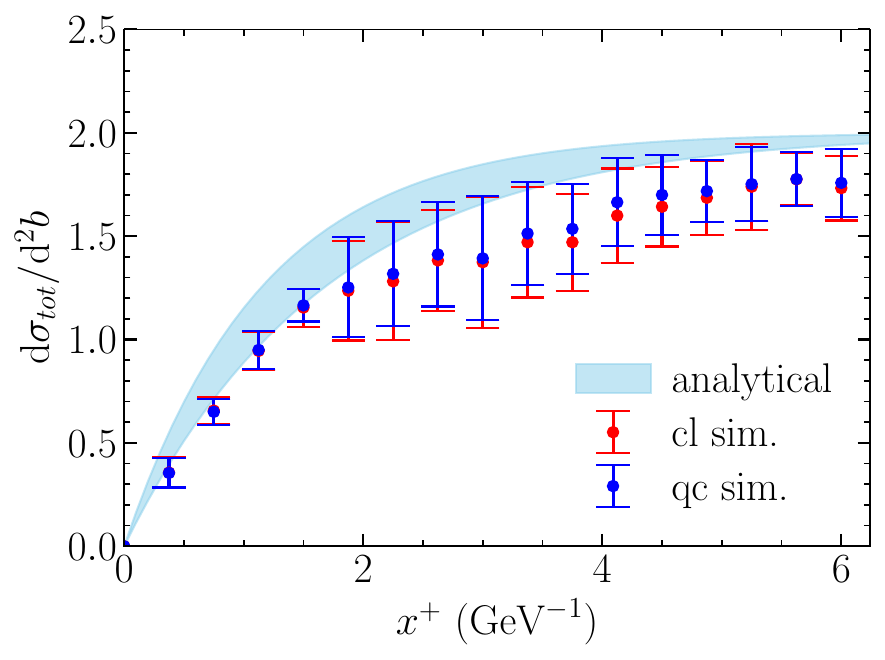}\quad\quad
    \includegraphics[width=0.45\linewidth]{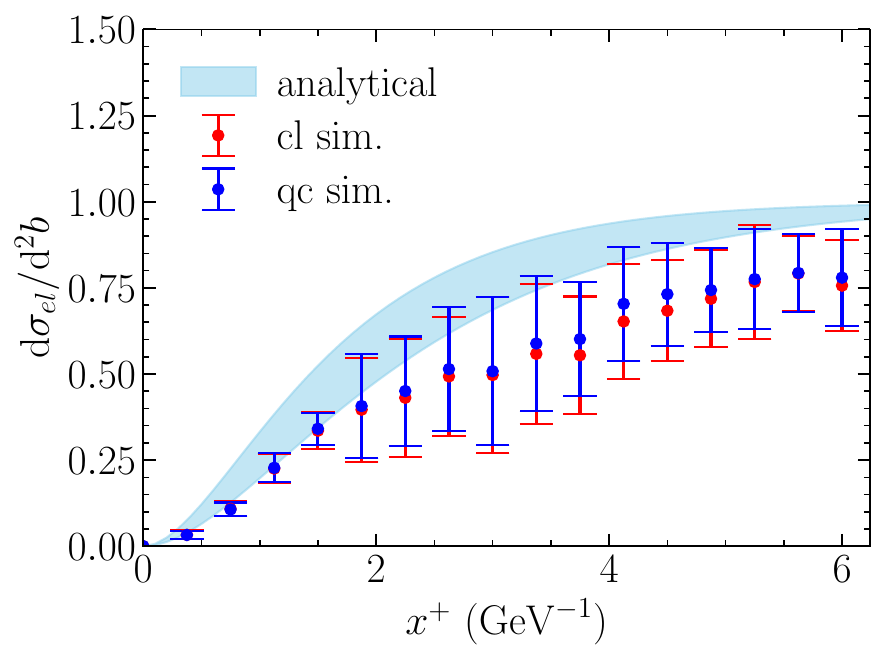}
    \caption{Total and elastic cross sections of single momentum dipole computed using classical and quantum circuit simulations.}
    \label{fig:xsection_qc}
\end{figure}

We also obtain dipole cross section using quantum simulation, with the framework that we have formulated in \cref{sec:qs}. We present in \cref{fig:xsection_qc} the total and elastic cross sections for single-momentum dipole initial state. The results using classical simulation are shown in red bar, and that using quantum simulation are in blue bar. The agreement between the two are very good in both total and elastic cross sections. Additionally, we show analytical results in blue band, calculated according to \cref{eq:crss_eikonal_qqbar_total}. The upper bound is obtained with $r_\perp=\infty$, and the lower obtained with $r_\perp=L_\perp/2$.

\subsection{Hadron production}

\begin{figure}
    \centering
    \includegraphics[width=0.5\textwidth]{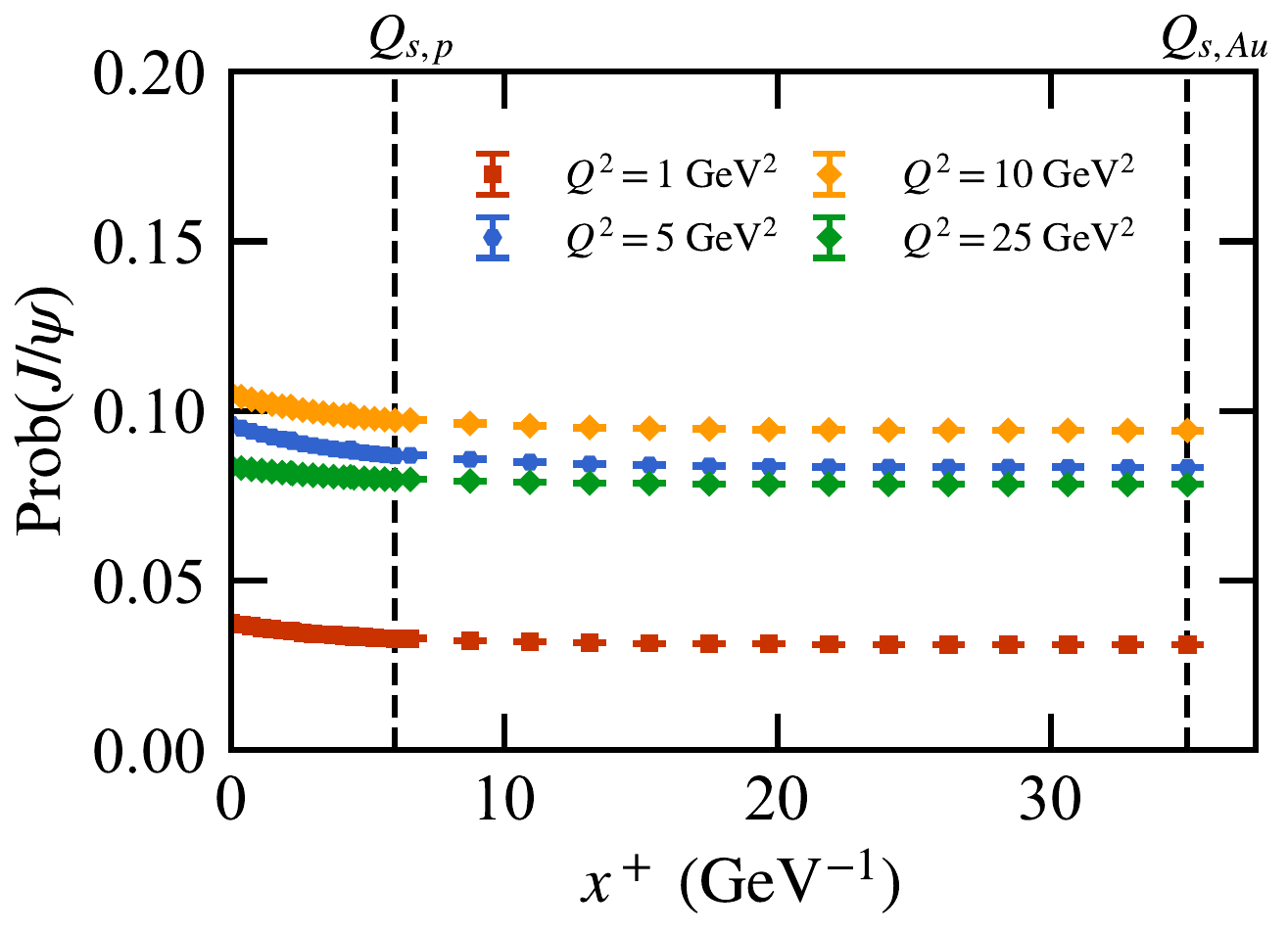}
    \caption{The $\gamma^*+\mathcal{A}\to J/\psi+X$ production probability as a function of the light-front time $x^+$ at various photon virtualities $Q^2$. }
    \label{fig:photon-Jpsi-production-t-Q2-proton}
\end{figure}

To understand the process of the quark-antiquark photon state evolving into the $J/\psi$ meson, we examine $\text{Prob}(J/\psi)$ as a function of evolution time.  For this exploratory study, we restrict our analysis to the $m_j=0$ component for the meson. In \cref{fig:photon-Jpsi-production-t-Q2-proton}, we present $\text{Prob}(J/\psi)$ as defined in \cref{eq:meson_prob} at various photon virtuality $Q^2$.
The production probability is maximal at early light-front times and decreases as the photon state propagates through the medium, reflecting that the interactions with the nuclear target modify the quark-antiquark configuration away from the $J/\psi$ state.
The dependence on $Q^2$ is non-monotonic: $\text{Pro}(J/\psi)$ increases with virtuality up to $Q^2\approx 10 \GeV^2$ and then decreases at higher $Q^2$. This behavior is consistent with the transverse momentum structure of the photon LFWF shown in \cref{fig:WF_compare}.
At lower $Q^2$, the photon LFWF grows in magnitude with increasing $Q^2$, enhancing its overlap with the $J/\psi$ wavefunction. At higher $Q^2$, however, the photon LFWF decreases in magnitude and broadens n $k_\perp$, reducing the overlap with the narrower $J/\psi$ wavefunction. As a result, the two wavefunctions have the greatest overlap at intermediate virtualities, yielding the largest production probability.

\begin{figure}[htp!]
    \centering
    \begin{subfigure}[b]{0.45\textwidth}
        \centering
        \includegraphics[width=\textwidth]{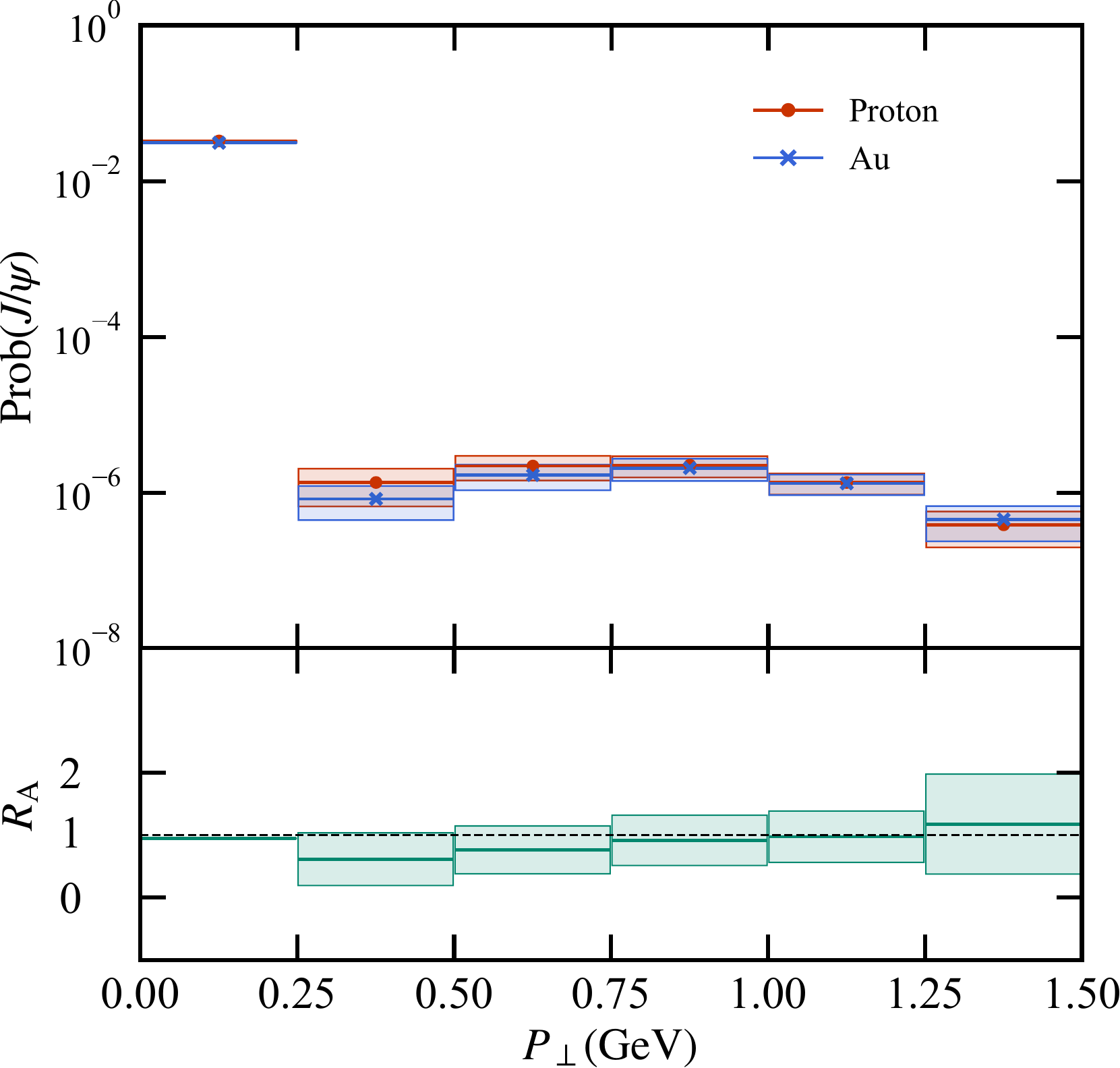}
        \caption{$Q^2 = 1~\text{GeV}^2$ \label{fig:photon-SU3-JPsi-production_Q2_1}}
    \end{subfigure}
    \hfill
    \begin{subfigure}[b]{0.45\textwidth}
        \centering
        \includegraphics[width=\textwidth]{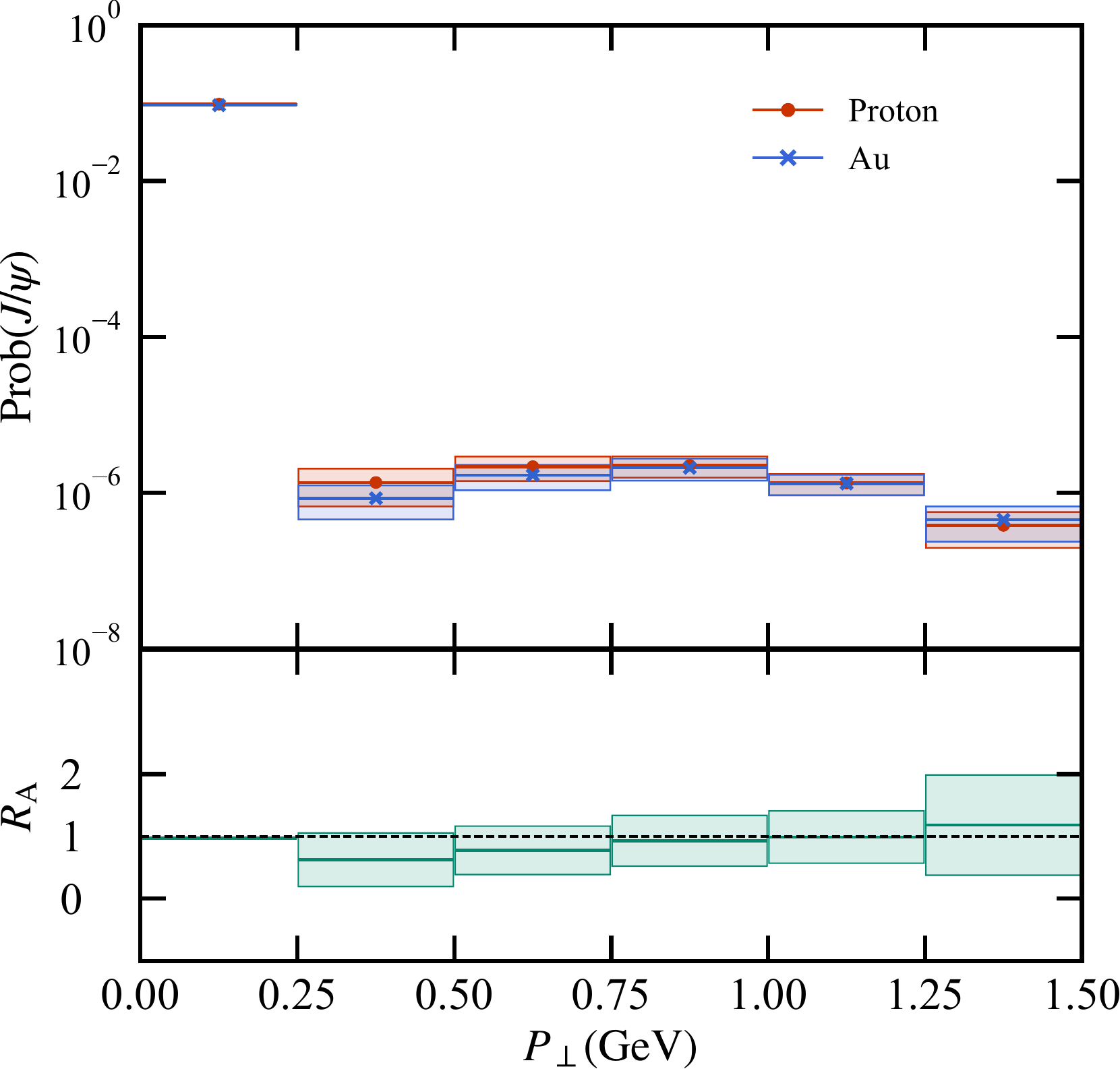}
        \caption{$Q^2 = 10~\text{GeV}^2$\label{fig:photon-SU3-JPsi-production_Q2_10}}
    \end{subfigure}
    \caption{The $J/\psi$ production probability is shown as a function of the $J/\psi$ center-of-mass transverse momentum. 
}
    \label{fig:photon-SU3-JPsi-production}
\end{figure}

In \cref{fig:photon-SU3-JPsi-production}, we present the $J/\psi$ production probability in terms of the C.M. transverse momentum $P_\perp$ for both the proton and the Au nucleus.
Here the sum over $\vec{P}_\perp$ in \cref{eq:meson_prob} is restricted to modes within each $P_\perp=|\vec{P}_\perp|$ bin, so the figure shows the production probability resolved by the magnitude of the C.M. transverse momentum.
Results are shown for photon virtualities $Q^2 = 1 \GeV^2$ in \cref{fig:photon-SU3-JPsi-production_Q2_1} and $Q^2 = 10 \GeV^2$ in \cref{fig:photon-SU3-JPsi-production_Q2_10}. 
In both cases, the production is dominated by the lowest bin, $P_\perp<0.25~\GeV$, which exceeds the others by about four orders of magnitude. 
Since the initial photon has $\vec P_\perp=0$, this shows that the C.M. momentum of the $q\bar q$ pair barely changes during the evolution.
As discussed for \cref{fig:photon-rel-t}, the quark and the antiquark in the photon are strongly correlated in transverse space, with a separation below the resolution of the basis. They therefore experience nearly the same background field, and the momentum transfers to the color-singlet pair largely cancel. 
Producing a $J/\psi$ at finite $P_\perp$ requires the pair to acquire a net transverse momentum, which only the small component with separated quark and antiquark can provide. Above $P_\perp=0.25~\GeV$, the probability is of order $10^{-6}$, varies only mildly, and decreases in the highest bin.
 
The nuclear modification factor $R_A$, shown in the lower panels, is consistent with unity in the lowest $P_\perp$ bin, reflecting that the C.M. momentum of the pair is largely unaffected by the target. 
In the region $P_\perp\sim0.25$-$0.5~\GeV$, $R_A$ indicates a suppression, and at larger $P_\perp$ it is consistent with unity within the uncertainties arising from fluctuations of the nuclear configuration.
The pattern is qualitatively similar for both values of $Q^2$.
The production probability in the lowest bin is larger at $Q^2 = 10 \GeV^2$ than at $Q^2 = 1 \GeV^2$. This is consistent with the $Q^2$ dependence observed in \cref{fig:photon-Jpsi-production-t-Q2-proton} and the discussion therein. Nuclear modification of $J/\psi$ production in photon-nucleus scattering has been studied within the CGC framework~\cite{Lappi:2010dd,Cheung:2024qvw}, though with different hadronization prescriptions.

\subsection{Quantum entanglement}
The entanglement entropy could be a useful indication of the status of the quantum many-body system independent of specific observables, which are seen in studies of bound-state problems between partons~\cite{Zhang:2025ean,Zhang:2026pff} and real-time dynamics~\cite{Barata:2023clv, Florio:2025hoc,Florio:2024aix,Florio:2023dke,Rigobello:2021fxw}.
Here, we compute the von Neumann (vN) entropy of the reduced density matrix obtained by tracing out the antiquark degrees of freedom,
\begin{align}\label{eq:entropy}
    S_{\mathrm{vN}}(x^+) = - \mathrm{Tr}[\bar\rho_q(x^+) \log_2 \bar\rho_q(x^+)]\,.
\end{align}
For each event, the density matrix is given by $\rho(x^+)  =\ket{\psi(x^+)}\bra{\psi(x^+)} $, with the state normalized to unity, and the reduced quark density matrix is $\rho_q(x^+)  =\Tr_{\bar q} \rho(x^+) $.
We then average over the $N_\mathrm{event}$ background-field configurations to get $\bar\rho_q =\braket{\rho_q}_{\text{event}}$. A similar density-matrix average
was used in Ref.~\cite{Barata:2023clv}, where the gluon is traced out in a
$\ket{q}+\ket{qg}$ system instead.

It is important to distinguish this entropy from the entropy of the reduced density matrix of a single event, $S_{\rm vN}[\rho_q(x^+)]$.
In a fixed background field $\mathcal A$, the quark and the antiquark interact with the field separately, and the evolution of the pair is a product of a unitary acting on the quark, $U_q(\mathcal A)$, and one acting on the antiquark, $U_{\bar q}(\mathcal A)$. 
The antiquark unitary drops out of the partial trace, so that $\rho_q(x^+)=U_q(\mathcal A)\,\rho_q(0)\,U_q^\dagger(\mathcal A)$.
The entropy $S_{\rm vN}[\rho_q(x^+)]$ is therefore constant in $x^+$.
The medium only rotates the basis in which the entanglement is encoded, through color rotations and transverse-momentum kicks.
The field $\mathcal A$ differs from event to event, so $U_q(\mathcal A)$ differs as well, and $\bar\rho_q$ is an ensemble average of differently rotated density matrices. 
By the concavity of the entropy~\cite{nielsen_chuang_2010},
\begin{align}
    S_{\rm vN}[\bar\rho_q(x^+)] \ge
    \braket{S_{\rm vN}[\rho_q(x^+)]}_{\text{event}}
    = S_{\rm vN}[\rho_q(0)]\,.
\end{align}
The physical meaning of $S_{\rm vN}[\bar\rho_q]$ therefore has two sources. 
The first is the entanglement with the antiquark, which is the same in every event and equals its initial value $S_{\rm vN}(0)=S_{\rm vN}[\rho_q(0)]$.
The second is the event-to-event fluctuation of the quark color and transverse momentum induced by the medium.
The growth of $S_{\rm vN}[\bar\rho_q]$ with $x^+$ thus does not mean that more entanglement is generated between the quark and the antiquark, but that the quark color and transverse momentum are spread more widely over the events. 
The maximal value of $S_{\rm vN}[\bar\rho_q]$ is $\log_2 n$, $n$ being the dimension of the Hilbert space of the quark.

\begin{figure}[htp!]
    \centering
    \begin{subfigure}[b]{0.45\textwidth}
        \centering
        \includegraphics[width=\textwidth]{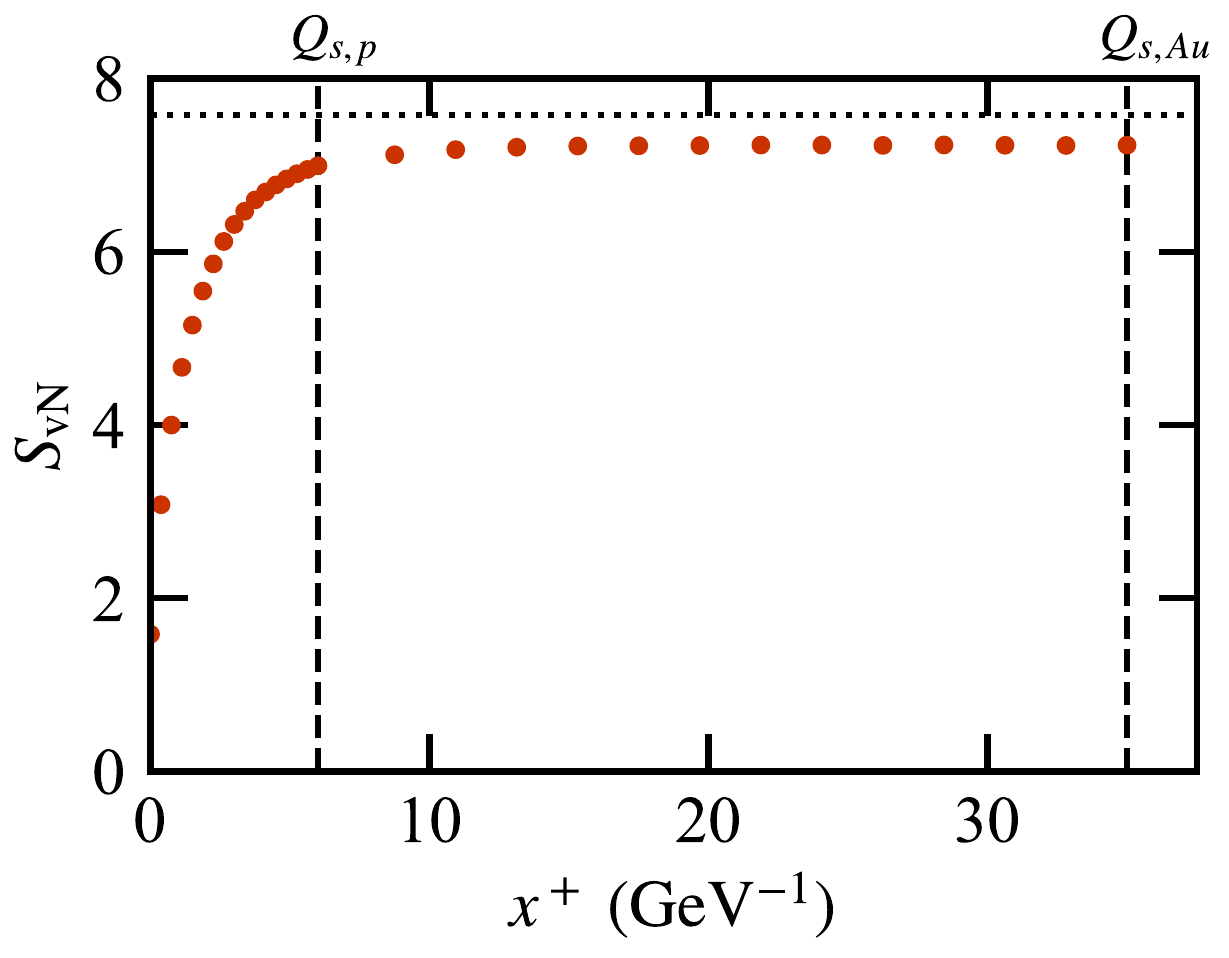}
        \caption{The single-momentum dipole  \label{fig:entropy_qqbar}}  
    
    \end{subfigure}
    \begin{subfigure}[b]{0.48\textwidth}
        \centering
        \includegraphics[width=\textwidth]{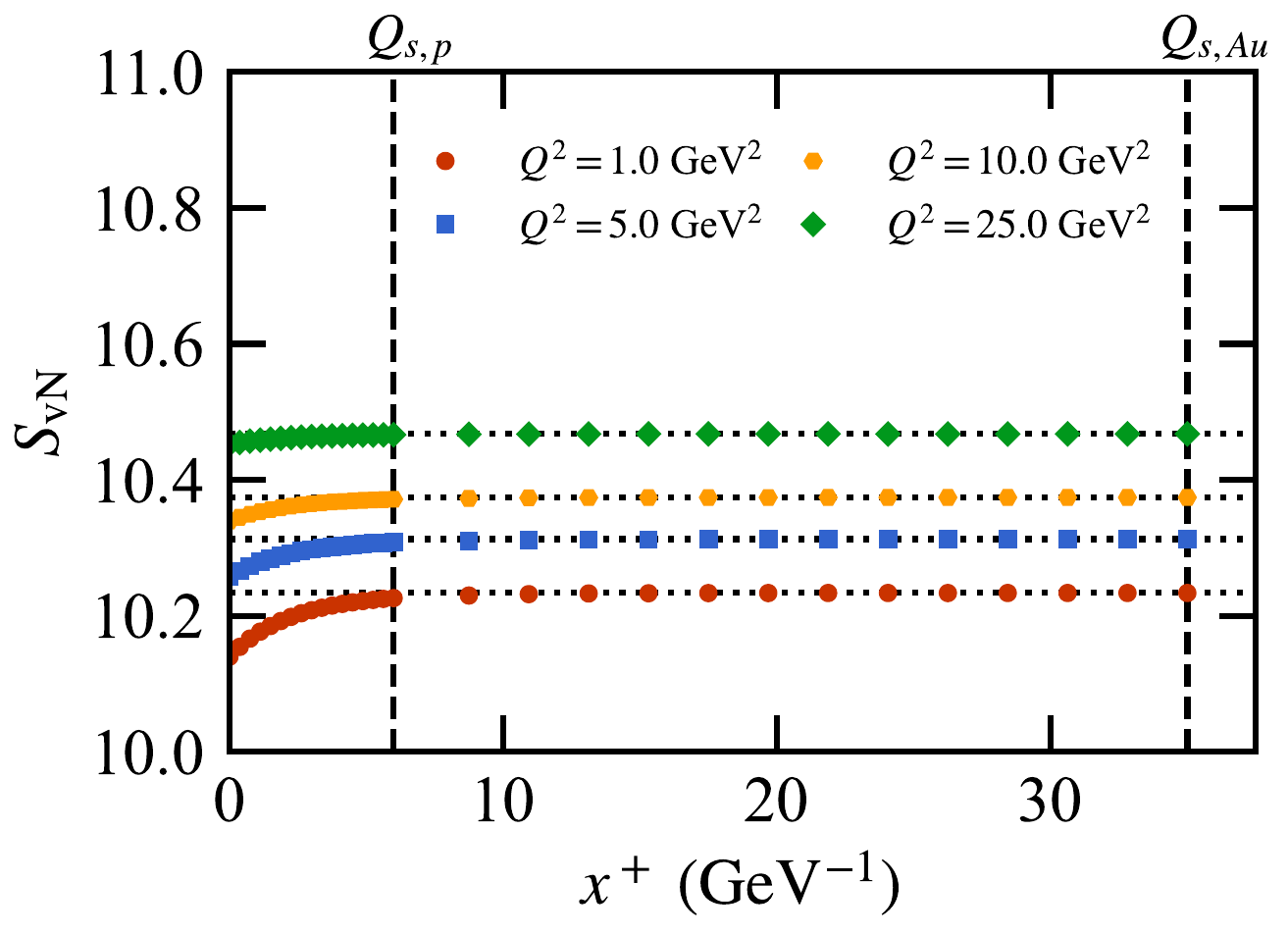}
        \caption{The virtual photon  \label{fig:entropy_photon}}  
    \end{subfigure}
    \caption{
    The von Neumann entanglement entropy, as defined in \cref{eq:entropy},  of the quark in the $|q\bar{q}\rangle$ Fock state for the both initial states. At the initial time, $S_{vN} = \log_2 N_c$ for the single-momentum dipole, reflecting only the color entanglement. Various virtualities $Q^2$ are shown for the photon state.
    \label{fig:entropy}
    }
\end{figure}

\Cref{fig:entropy} shows $S_{\rm vN}[\bar\rho_q]$ for the probe evolving in the nuclear targets.
For the single-momentum dipole, shwon in \cref{fig:entropy_qqbar}, $S_\mathrm{vN}$ starts at $\log_2 N_c\approx 1.6$, the entanglement of the initial color-singlet state. It grows rapidly as the event average mixes the color and transverse-momentum configurations, and levels off near $x^+\sim 6 \GeV^{-1}$ once the distribution has spread over the available phase space.
The dotted line marks the upper bound $\log_2[N_c(2N_\perp)^2]\approx 7.6$, which corresponds to a quark completely randomized over the $N_c(2N_\perp)^2$ modes in color and transverse momentum.
At late times the entropy is close to this bound at $\approx 7.2$, but slightly below it, so the distribution is close to but not completely uniform.

For the photon state, shown in \cref{fig:entropy_photon}, the initial entropy is already large because the photon wavefunction is a superposition of many color, transverse-momentum and longitudinal-momentum modes that are entangled between $q$ and $\bar q$. 
The subsequent growth is therefore modest, and saturation occurs on about the same time scale as for the single-momentum dipole.
Counting the $K$ longitudinal-momentum modes, the naive upper bound is $\log_2[K N_c(2N_\perp)^2]\approx 10.6$.
It is not attained because the medium transfers no $p^+$ to the dipole. 
The probability distribution $P(x)$ of the quark longitudinal momentum fraction $x$ is given by the photon wavefunction in \cref{eq:photon_LFWF} and does not change during the evolution.
Consequently, the entropy can be at most $H_x+\log_2[N_c(2N_\perp)^2]$, where $H_x=-\sum_x P(x)\log_2 P(x)$ is the Shannon entropy of this distribution, and $H_x<\log_2 K$ unless $P(x)$ is uniform.
This bound depends on $Q^2$ through $P(x)$ and is shown by the dotted line for each $Q^2$ in \cref{fig:entropy_photon}.
The entropy saturates close to the corresponding bound, so the photon states with different virtualities saturate to different values.

\section{Summary and discussions \label{sec:summary}}
We have presented a first-principles quantum simulation framework for studying meson production in deep inelastic scattering. Using a light-front QCD Hamiltonian formalism, we evolved a $q\bar{q}$ dipole through a nuclear color field and
extracted physical observables such as transverse momentum, cross sections, hadron production rate, and nuclear  modification factor from the resulting quantum state with both classical simulation and quantum-circuit simulation. Our results are compared to analytical expectations when possible and show reasonable agreement despite being simulated on a small basis space.

We first studied the transverse-momentum dynamics of the dipole. The quark momentum broadening grows linearly at early times, following the analytical prediction, and saturates at late times once the distribution has spread over the finite set of transverse modes. The quantum-circuit results agree with the classical benchmark within event uncertainties. For the photon initial state, the center-of-mass momentum of the $q\bar q$ pair remains near zero throughout the evolution, reflecting the color transparency of the compact dipole on the lattice.

We then computed the total and elastic dipole cross sections as functions of the light-front time. 
For the single-momentum dipole, the results agree with the analytical eikonal predictions and approach the black-disk limit for the Au targets. 
For the photon state, the approach to saturation is slower, and the cross sections exhibit a non-monotonic dependence on the photon virtuality $Q^2$, reflecting the interplay between the varying amplitude of the $q\bar q$ component and color transparency.

The $J/\psi$ production probability was obtained by projecting the evolved photon state onto the $J/\psi$ LFWF. The production probability is concentrated at $P_\perp\approx 0$, consistent with color transparency of the compact $c\bar c$ pair. 
The nuclear modification factor $R_A$ is consistent with unity at $P_\perp\approx 0$ and shows suppression at small $P_\perp$.
We also studied the von Neumann entropy of the event-averaged reduced quark density matrix.
Its growth reflects the spread of the quark color and transverse momentum induced by the medium, and it saturates near the value of a fully randomized quark.
For the photon state, the saturation value depends on $Q^2$, because the longitudinal momentum distribution is not changed by the medium.
The entropy may thus serve as a diagnostic of medium-induced decoherence.

Several directions remain for future work: first, the preparation of the photon initial state on the circuit, which requires a general $W$ state preparation beyond the zero-momentum singlet considered, is an important next step toward full EIC-relevant kinematics.
Second, the current implementation works in the eikonal limit $p^+\to\infty$; extending this approximation to include finite-$x^+$ kinematic and non-eikonal corrections will require incorporating the kinetic energy term $P^-_{\rm KE}$ in the evolution, which is straightforward within the present encoding but increases the circuit depth. 
Third, going beyond the $\ket{q\bar{q}}$ Fock sector to include gluon degrees of freedom is the natural extension that would capture gluon radiation and Fock-sector mixing during propagation. 
Finally, scaling to larger transverse lattices and $N_c=3$ on quantum hardware will benefit from error mitigation strategies and the Q-SOFT approach~\cite{Barata:2023clv}, which reduces the gate count by exploiting the diagonal structure of the kinetic and interaction terms in momentum and position space, respectively.

\subsection*{Acknowledgements}

X.-h. Cao is supported by the National Natural Science Foundation of China (NSFC) under Grant No.~125B2111.
M. Li is supported by new faculty startup funding from Huazhong University of Science and Technology.
Y. Li is supported by the Chinese Academy of Sciences under Grant No. YSBR-101.
W. Qian is supported by the NSFC under Grant No. 12675177.
M. Li and W. Qian are also supported by IGNITE 2025, funded by Maria de Maeztu Excellence Unit budget, project code CEX2023-001318-M of the Agencia Estatal de investigación (AEl, Spanish Agency for Research).

\bibliography{main.bib}

\end{document}